\documentclass[screen,authorversion,oneside]{acmart}
\AtBeginDocument{%
  }

\usepackage{booktabs}
\usepackage{colortbl}
\usepackage{listings}
\usepackage{longtable}
\usepackage{placeins}
\usepackage{xcolor}

\definecolor{codegray}{rgb}{0.95,0.95,0.95}

\begin{document}

\title[Connecting the Dots]{Connecting the Dots: \\ Surfacing Structure in Documents through AI-Generated Cross-Modal Links}


\author{Alyssa Hwang}
\email{alyssahwang@alumni.upenn.edu}
\orcid{0009-0006-4827-8505}
\affiliation{%
  \institution{University of Pennsylvania}
  \city{Philadelphia}
  \state{PA}
  \country{USA}
}

\author{Hita Kambhamettu}
\email{hitakam@seas.upenn.edu}
\orcid{0000-0001-9620-1533}
\affiliation{%
  \institution{University of Pennsylvania}
  \city{Philadelphia}
  \state{Pennsylvania}
  \country{USA}
}

\author{Yue Yang}
\email{yueyang1@seas.upenn.edu}
\affiliation{%
  \institution{University of Pennsylvania}
  \city{Philadelphia}
  \state{Pennsylvania}
  \country{USA}
}

\author{Ajay Patel}
\email{ajayp@seas.upenn.edu}
\affiliation{%
  \institution{University of Pennsylvania}
  \city{Philadelphia}
  \state{Pennsylvania}
  \country{USA}
}

\author{Joseph Chee Chang}
\email{josephc@allenai.org}
\orcid{0000-0002-0798-4351}
\affiliation{%
  \institution{Allen Institute for AI}
  \city{Seattle}
  \state{Washington}
  \country{USA}
}

\author{Andrew Head}
\email{head@seas.upenn.edu}
\orcid{0000-0002-1523-3347}
\affiliation{%
  \institution{University of Pennsylvania}
  \city{Philadelphia}
  \state{Pennsylvania}
  \country{USA}
}

\renewcommand{\shortauthors}{Hwang et al.}

\begin{abstract}
Understanding information-dense documents like recipes and scientific papers requires readers to find, interpret, and connect details scattered across text, figures, tables, and other visual elements. These documents are often long and filled with specialized terminology, hindering the ability to locate relevant information or piece together related ideas. Existing tools offer limited support for synthesizing information across media types. As a result, understanding complex material remains cognitively demanding. This paper presents a framework for fine-grained integration of information in complex documents. We instantiate the framework in an augmented reading interface, which populates a scientific paper with clickable points on figures, interactive highlights in the body text, and a persistent reference panel for accessing consolidated details without manual scrolling. In a controlled between-subjects study, we find that participants who read the paper with our tool achieved significantly higher scores on a reading quiz without evidence of increased time to completion or cognitive load. Fine-grained integration provides a systematic way of revealing relationships within a document, supporting engagement with complex, information-dense materials.
\end{abstract}

\ccsdesc[100]{Human-centered computing~User studies}
\ccsdesc[300]{Human-centered computing~Graphical user interfaces}
\ccsdesc[500]{Human-centered computing~Hypertext / hypermedia}

\keywords{fine-grained integration, augmented reading interfaces, AI interfaces, AI augmentations}
\begin{teaserfigure}
  \includegraphics[width=\textwidth]{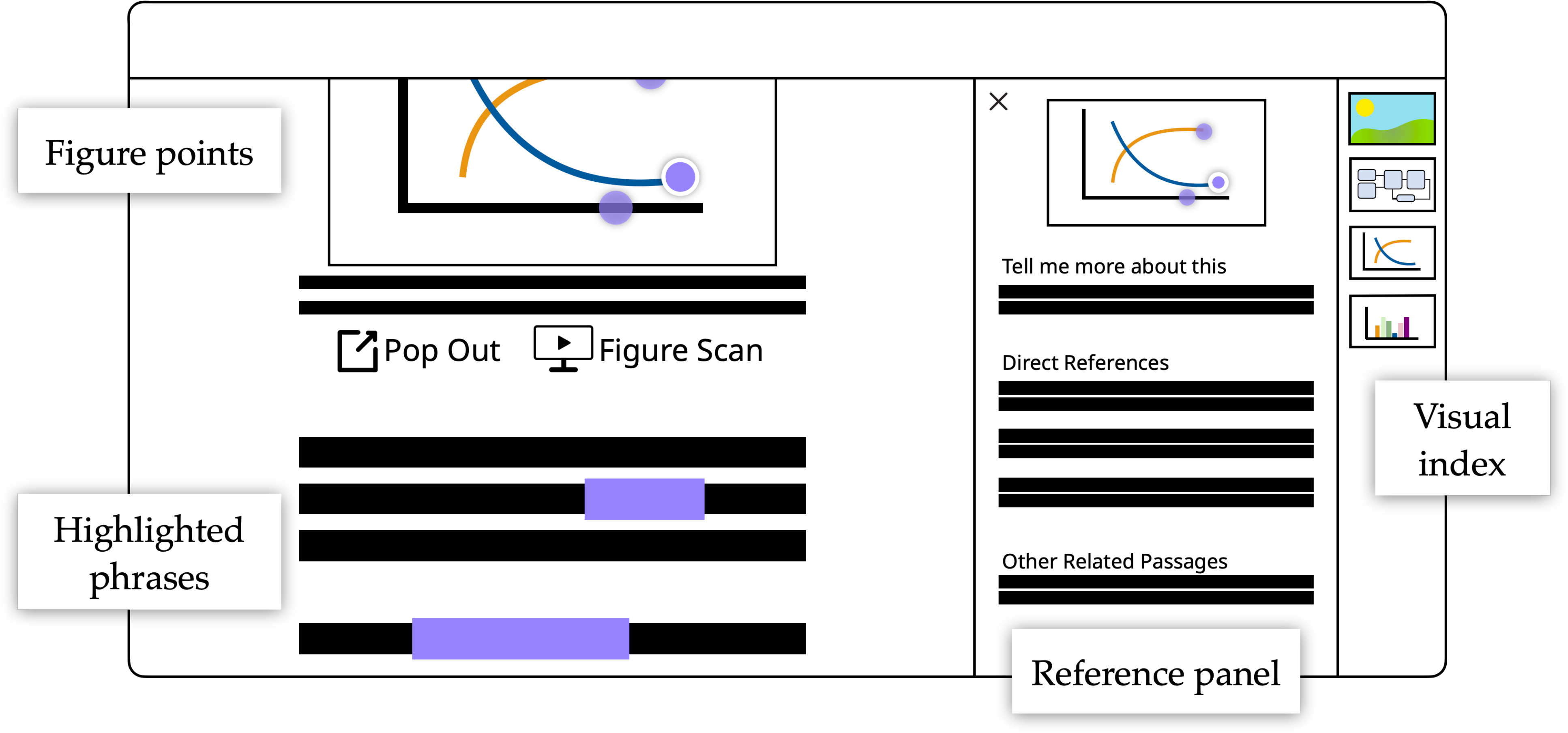}
  \caption{Fine-grained integration at document scale. We present an abstract framework for integrating information into complex documents at fine granularities, which is instantiated as a reading interface for research papers. The framework identifies entities in figures (``figure points'') and text passages (``highlighted phrases'') consolidates linked details in the reference panel, and allows users to navigate by clicking on figures in the visual index (see Sections \ref{sec:design} and \ref{sec:system}). The framework was developed with feedback from users (Section \ref{sec:formativestudy}) and evaluated with a between-subjects user study (Section \ref{sec:eval}).}
  \Description{Abstract visualization of our interface with labeled components (``Figure points,'' ``Highlighted phrases,'' ``Reference panel,'' and ``Visual index.''}
  \label{fig:teaser}
\end{teaserfigure}

\received{17 February 2026}

\maketitle

\section{Introduction}
\label{sec:introduction}

Many contemporary documents are long, information-dense, and multimodal, combining text with figures, tables, and equations. In such documents, information relevant to a single concept is often distributed across sections and modalities, requiring readers to locate, connect, and integrate scattered details. Understanding may depend on background knowledge that is not explicitly provided, forcing readers to infer missing connections or consult external references. As a result, even experienced readers can find these documents cognitively demanding and time-consuming to navigate.

Prior work has made substantial progress in supporting comprehension of textual content, for example through summarization tools, highlighting interfaces, and reading aids that clarify difficult passages. These approaches primarily support understanding at the level of sections or documents as a whole. Other kinds of tools elaborate on individual artifacts, like annotating or animating one figure at a time. These tools provide limited support for fine-grained, cross-media reading at document scale. This limitation is especially pronounced in multimedia documents like scientific papers, where critical details may appear in figures, captions, tables, or appendices along with the main text.

Supporting these fine-grained reading tasks requires new ways of representing and surfacing relationships across modalities, as well as interaction techniques that allow readers to efficiently access and integrate related information. Recent advances in multimodal large language models (MLLMs) make addressing these challenges increasingly feasible. Models such as GPT-4o~\citep{openai_gpt-4o_2024} and Molmo~\citep{deitke_molmo_2024} can reason about visual content and identify specific elements within images, enabling fine-grained links between text and visuals that previously required domain-specific heuristics~\citep{huang_detection_2024}. While many systems now demonstrate the technical feasibility of such links, there is limited guidance on how they should be designed and integrated into interactive reading experiences. Designing such systems raises open questions about how connections should be structured, when they should be revealed, and how they can support readers without disrupting their reading flow.

In this paper, we propose a framework for fine-grained integration that defines design principles for exposing relationships among small, distributed details across text and visuals. The framework provides a conceptual lens for understanding how AI-driven systems can support readers in connecting related information in ways that align with human reading and sensemaking practices. We ground this work in the domain of scientific papers, which are characterized by long, multimodal documents with complex figures~\citep{lee2017viziometrics, cordero2016life}.

The development of our framework is guided by the following research questions:

\begin{itemize}
\item \textbf{RQ1 (Reading Challenges)}: What challenges do readers face when interpreting and connecting visual and textual information in long, multimedia documents?
\item \textbf{RQ2 (Design Opportunities)}: How can fine-grained integration be designed to clarify details and reveal relationships across a document?
\item \textbf{RQ3 (Framework Development)}: What principles and affordances, derived from iterative design and user feedback, can inform a generalizable framework for fine-grained integration?
\end{itemize}

This paper contributes (1) a framework for fine-grained integration of related information within information-dense documents, (2) empirical grounding for this framework through formative and think-aloud studies examining how readers explore, connect, and interpret information when engaging with augmented reading materials, and (3) findings from a controlled between-subjects study that demonstrate statistically significant improvement in reading quiz performance without evidence of increased time or cognitive load when reading with our interface.

\section{Related Work}
\label{sec:rw}

\textit{Cross-modal links.} A growing body of work has explored how computational tools can help readers integrate textual and visual information more effectively. One central strategy is to link parts of body passages to related elements across ``floating'' document components, like tables \citep{badam_elastic_2019, kim_facilitating-document-reading_2018, zhu-tian_crossdata_2022}, charts \citep{latif_kori_2022, kong_extracting_2014}, and videos \citep{kim_papeos_2023}. Many approaches to cross-modal links are component-centric; for example, they focus on expanding a user's understanding of a particular paragraph or figure. Building on this line of work, we investigate how cross-modal links can be scaled up to support the understanding of many figures and entire documents simultaneously.

\textit{Augmented artifacts.} Beyond linking, recent work shows how visual artifacts themselves can serve as interactive entry points for understanding complex information. Many systems support ``details on demand'' \citep{shneiderman2003eyes}, such as providing textual explanations in response to natural-language questions about a chart \citep{kim_answering_2020} or revealing information about specific data points \citep{latif_exploring_2018}. Other approaches enable direct interaction with visual artifacts by highlighting selected regions \citep{kong_graphical_2012, ying_reviving_2024} or presenting step-by-step animations of diagrams or charts \citep{ying_reviving_2024, grossman_your_2015, gunturu_augmented_2024, lai_automatic_2020}. Related work on dynamic or ``live'' figures allows users to manipulate parameters or explore variations of the underlying data \citep{dragicevic_increasing_2019, heer_living_2023, chulpongsatorn_augmented_2023, gunturu_augmented_2024}. We extend this line of work to the document scale by enabling all figures in a paper to act as augmented artifacts, each supporting links to related text and step-by-step walkthroughs.

\textit{Attention cues.} Another line of work examines how attention cues can help readers manage information density in both visual and textual materials. In text, techniques such as color-coded highlights \citep{fok_scim_2023} and selective fading or deemphasis \citep{gu_ai-resilient_2024, head_augmenting_2021} can guide readers toward relevant passages while reducing distraction from other content. In visualizations, attention can be directed through animation, narration, or annotations that emphasize particular elements within complex figures \citep{lai_automatic_2020, latif_exploring_2018, sultanum_datatales_2023, ying_reviving_2024, kong_graphical_2012}. Our work integrates attention cues across both text and figures, enabling coordinated guidance that spans modalities within a single document.

\section{Formative Study}
\label{sec:formativestudy}

We begin with a formative study to examine reading challenges specific to information-dense documents. Using scientific papers as a representative setting, we study how readers make sense of information distributed across text and figures. This study establishes the foundation for our framework, grounding it in the real-world practices and difficulties of reading complex materials. This study was approved by the University of Pennsylvania Institutional Review Board.

\subsection{Participants}

\begin{table}[]
\centering
\begin{tabular}{llll}
\toprule
ID  & Occupation & Research Area                     & Reading Format   \\
\midrule
F1  & PhD        & experimental physics              & laptop + tablet  \\
F2  & PhD        & history of science and technology & laptop           \\
F3  & PhD        & operations                        & printed          \\
F4  & PhD        & operations                        & laptop + notepad \\
F5  & PhD        & computer science                  & tablet           \\
F6  & PhD        & computer science                  & laptop + notepad \\
F7  & PhD        & computer science                  & laptop           \\
F8  & PhD        & computer science                  & laptop           \\
F9  & PhD        & computer science                  & printed          \\
F10 & Postdoc    & anthropology                      & printed \\  
\bottomrule
\end{tabular}
\caption{Formative study participants. Participants represented a range of fields and reading preferences from the University of Pennsylvania research community.}
\label{tab:formative-participants}
\end{table}

We recruited ten researchers from the University of Pennsylvania to participate in the formative study (see Table \ref{tab:formative-participants}). With the exception of one postdoctoral researcher in anthropology, all participants were PhD students from a range of disciplines: experimental physics (1), history of science and technology (1), operations (2), computer science (4), and sociology (1). Participants were compensated with a \$30 USD virtual gift card.

\subsection{Procedure}
Each participant completed an individual one-hour session at the University of Pennsylvania Human-Computer Interaction Lab. Before their sessions, participants were asked to bring a research paper of their choice in whichever format they preferred, including laptops, printed copies, or tablets (see Table \ref{tab:formative-participants}). To ensure sufficient visual content, we requested that participants select papers containing multiple figures.

The facilitator started each session by introducing the study and reviewing the consent form. The main activity was an independent reading session, during which participants were instructed to read as they usually would while voicing their thoughts aloud. The facilitator observed, took notes on a laptop, asked occasional clarifying questions, and recorded audio for later analysis.

At the end of each session, the facilitator conducted a semi-structured interview to learn more about the participant's reading process. Participants were asked about their usual reading strategies, aspects of reading that were easiest or most difficult, types of tools that they might find useful, and challenges they typically encounter when reading research papers. The session then concluded with a demographics questionnaire.

\subsection{Reading Challenges}
\label{sec:reading-challenges}

We conducted a thematic analysis of the session notes following established qualitative analysis procedures~\citep[Chapter 5]{blandford_qualitative_2016}. Our analysis identified three primary challenges associated with reading research papers. We refer to formative study participants through anonymous identifiers F1--F10.

\smallskip

\textit{Fragmentation.}
Participants emphasized that figures were essential for understanding a paper but difficult to connect with related text. As F2 noted, ``figures are important because they often contain clarifying examples,'' but they ``cannot be read alone because the prose indicates why they matter.'' Readers frequently encountered difficulties when information was fragmented across sections, such as when figures were placed far from their references or located in appendices (F1, F3, F5). F3 expressed frustration when two figures depicted related ideas without clear connections, while F2 struggled when a single figure was split across distant sections, prompting repeated scrolling. Participants also reported that uninformative captions and missing explanatory details further hindered comprehension (F3, F6, F8).

\textit{Complexity.}
Most participants (8/10) reported that figure complexity posed a major challenge when reading research papers. Several, including F2 and F6, expressed frustration with figures that were difficult to interpret or overly dense. As F2 explained, readers often had ``no base case to fall back on'': if the figure was unclear, no alternative explanation was available. In some fields, such as experimental physics, complex figures were expected; F9 noted that such figures assumed substantial domain expertise and offered little background information. Participants described being overwhelmed by figures with many components (F3, F8), particularly when faced with unclear notation (F1, F2, F6), ambiguous labels (F6, F10), confusing color schemes (F6), or small, intricate details (F8). They emphasized a need for clearer explanations, step-by-step guidance, and tools that could break figures and related text into smaller, more manageable parts.

\textit{Interpretation.}
Participants also described difficulty understanding what figures were meant to convey, even when the visual elements themselves were clear enough to parse. Several readers, including F2 and F6, wanted some way to confirm that their interpretation matched what the authors intended. F2 and F3 both mentioned uncertainty about basic evaluative cues, such as which side of a graph represented improvement or which system was performing better, which often pushed them back into the main text to look for explicit claims. Some participants preferred reading explanations over inferring meaning from visuals (F7, F8), and many asked for more concrete support: definitions of variables, brief summaries of components, or short examples that clarified what the figure was actually doing (F6). Even readers who believed figures should stand alone found themselves stumbling over unclear labels or ambiguous structure (F10). These experiences pointed to a consistent challenge in reading figures: recognizing not only how to decode them but how to understand the specific point they are trying to make.

\begin{figure}[H]
    \centering
    \includegraphics[width=0.6\linewidth]{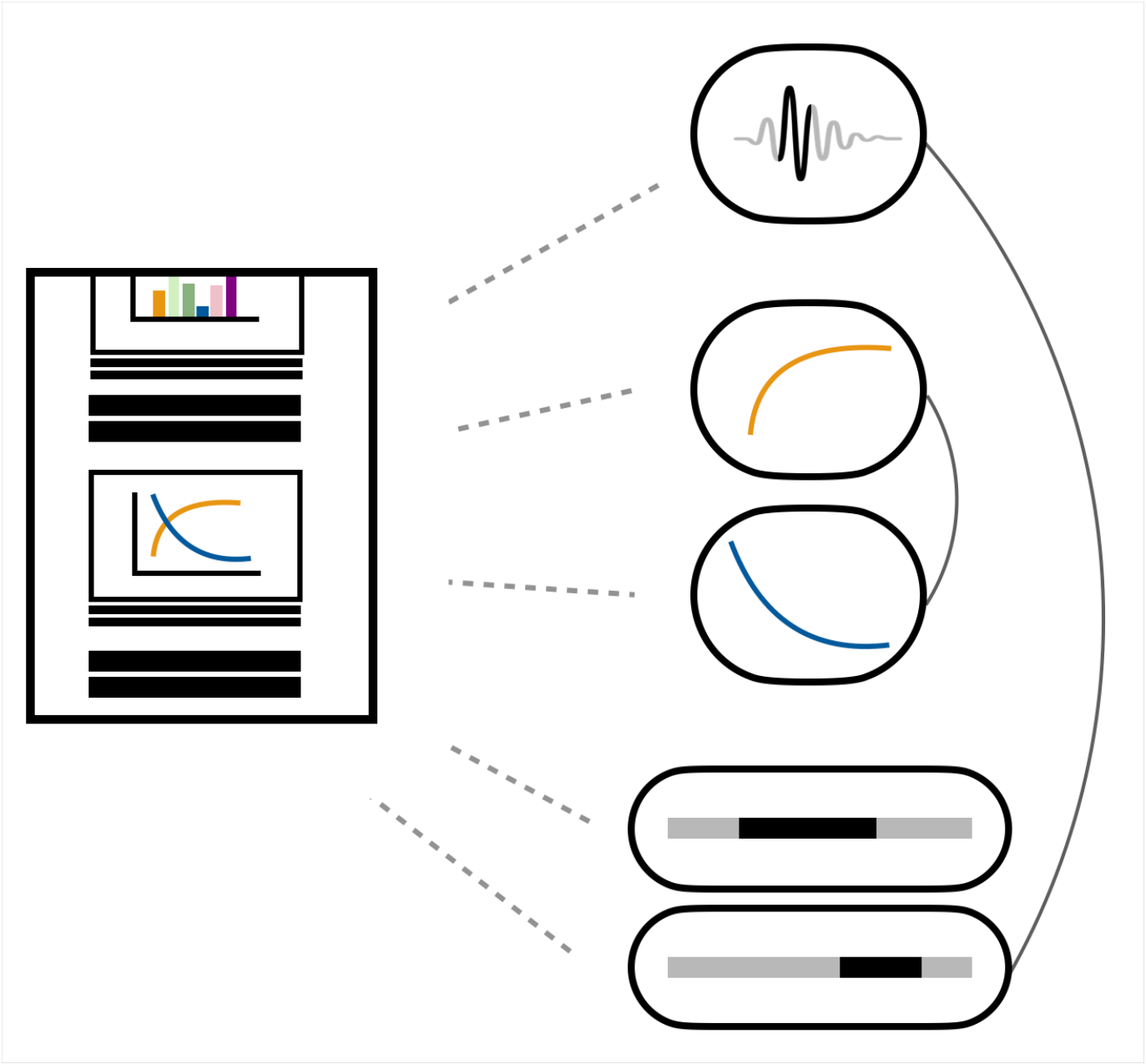}
    \Description{Diagram of our framework.}
    \caption{Framework design. Our framework for fine-grained integration consists of \textbf{entities} (segments in ovals) and \textbf{links} (gray curves between entities). See Section \ref{sec:design} for more details. ``Music'' image from Indygo at flaticons.com.}
    \label{fig:framework}
\end{figure}

\section{Framework for Fine-grained Integration}
\label{sec:design}

During our formative study, we observed key challenges related to the fragmentation of information throughout a long, complex document (see Section \ref{sec:reading-challenges}). Relevant details may be located paragraphs or even pages away from each other, requiring users to skim, scroll, or otherwise search the document when they need additional context. The process can be further complicated when key details are distributed across modalities, such as in body passages and figure images. Fragmentation can inhibit the reading process by making follow-up information more difficult to find, leading to lower understanding or slower navigation.

To address these challenges, we present a framework for integrating information in complex documents. Our framework focuses on integrating small details at fine granularities, like phrases in a paragraph or elements in an image. The framework is intended to include any combination of modalities, such as text, image, video, audio, or AR/VR.

\smallskip

\textit{Entities.} Our framework consists of two basic units: entities and links. An \textbf{entity} represents any discrete, interpretable, semantically meaningful item within a document. Abstractly, an entity may be a claim, piece of evidence, definition of a term, sample of data, or an important idea. Entities may be instantiated as text phrases, video clips, image segments, or other concrete artifacts. In general, entities are the pieces of information that users are searching for and consuming when reading, watching, or listening to complex information.

\textit{Links.} The second unit of our framework is a \textbf{link}, which represents a relationship between two entities. Two entities may be linked because one explains, defines, supports, contradicts, or provides an example for the other, among other possibilities. Links and entities can appear in documents in many different ways, like a paragraph explaining an element of a diagram or a video clip illustrating a concrete example of a new theory.

\smallskip

We developed our framework to be minimalist and abstract, which allows it to manifest in many ways depending on the type of information and the modality of interaction. In Section \ref{sec:system}, we instantiate our framework as a reading tool for figure-heavy research papers, demonstrating how entities and links can be exposed to users in the text and image domains. We evaluate the implications of the framework and interface in a controlled between-subjects user study, which we discuss in Section \ref{sec:eval}.

\section{System}
\label{sec:system}

\begin{figure}
    \centering
    \includegraphics[width=\linewidth]{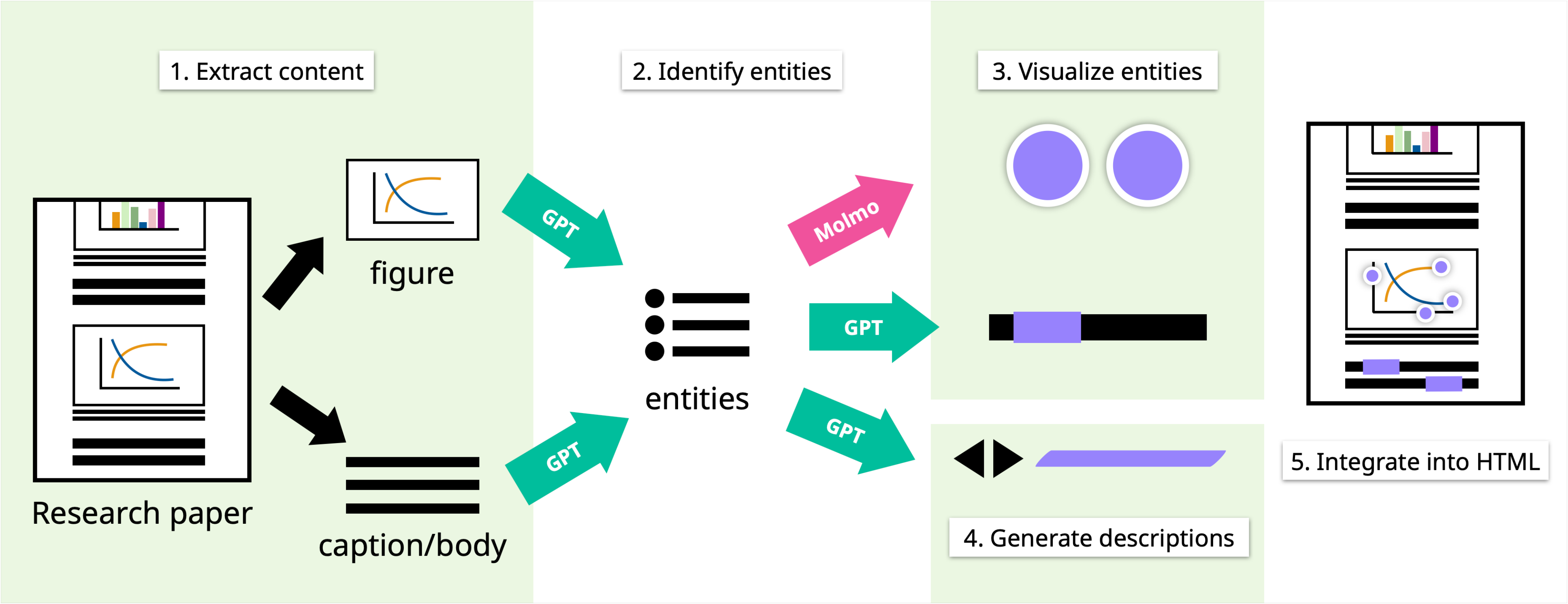}
    \caption{AI data generation pipeline. To generate data for our interface, we extract all figure images, captions, and referring passages. A multimodal OpenAI model identifies salient visual entities in figures and corresponding references in text, which are visualized as purple circles and highlighted phrases. In a separate pass, we provide the full paper as input to generate descriptions for all of the entities. The visual entities, textual references, and descriptions are integrated into the interface as interactive points, highlights, figure scans, and other affordances. Additional details are provided in Section \ref{sec:system}.}
    \Description{Five-part diagram of data generation process: (1) extract content, (2) identify entities, (3) visualize entities, (4) generate descriptions, (5) integrate into HTML.}
    \label{fig:pipeline}
\end{figure}

We instantiate our framework (Section \ref{sec:design}) as a reading interface for research papers. Our interface was implemented as an extension of the ACM (Association for Computing Machinery) Digital Library. We augmented the HTML version of recent conference papers with JavaScript for the frontend and Python for the backend, including an AI data processing pipeline. In this section, we discuss the key system and implementation details, with additional implementation details included in Sections \ref{app:design_implementation} and \ref{app:design_prompts}.

\subsection{Entities and Links}
\label{sec:entitieslinks}

Entities appear in research papers as text phrases and portions of images. As shown in Figure \ref{fig:pipeline}, we extracted entities with a pipeline including OpenAI GPT models and Molmo (\texttt{allenai/Molmo-7B-D-0924} from Hugging Face). The original version of our interface used \texttt{gpt-4o-2024-08-06}, while the updated version described in Section \ref{sec:eval_probe} used \texttt{gpt-5-2025-08-07}.

\smallskip

\textit{Image processing.} We automatically extracted entities from figures by prompting the OpenAI model to identify salient elements in the images (see Section \ref{app:design_prompts_image} for the full prompt). We then prompted Molmo one entity at a time to identify coordinate locations of the entities in the figure images (see Section \ref{app:design_prompts_coordinates} for the full prompt). We used this two-part approach because neither model alone supported the data processing behavior we needed at the time. We annotated figures with interactive ``figure points'' at each coordinate to represent each visual entity.

\textit{Text processing.} We used the OpenAI model to identify references to those entities in figure captions and body passages, which eventually appeared as highlighted spans of text in the document. To avoid overwhelming users with visual clutter, we constrained the body passages to those directly referencing a particular figure, which we identified through a regular expression (e.g., for entities seen in Figure 3, we provided body passages with mentions like ``in Figure 3'' or ``see fig. 3''). We instructed the model to identify phrases corresponding to a particular entity, even if they were not an exact match. This allowed us to unify phrases that refer to the same entity in different ways (e.g., ``Step 1'' and ``the first step'') (see Section \ref{app:design_prompts_text} for the full prompt). The visual and text entities are linked as ``direct references'' to each other in the reference panel (see ``Consolidation'' in Section \ref{sec:system}).

\textit{Entity descriptions.} Along with extracting visual and text entities from the paper's body passages, captions, and figure images, we introduced machine-generated descriptions of each entity (see Section \ref{app:design_prompts_walkthroughs}). These descriptions were intended to provide an in-depth, standalone explanation of each entity without requiring the user to have read the rest of the paper beforehand. They may synthesize, summarize, or emphasize details found throughout the paper. The descriptions appear in our interface in two ways: in a consolidated reference panel (see ``Consolidation'' in Section \ref{sec:system}) and a ``figure scan'' (``Decomposition'' in Section \ref{sec:system}). We also instructed the model to identify related sentences from throughout the paper. In the interface, these sentences are linked to visual and text entities as ``other related passages.''

\subsection{Affordances}
\label{sec:system_affordances}

\begin{figure}
    \centering
    \includegraphics[width=\linewidth]{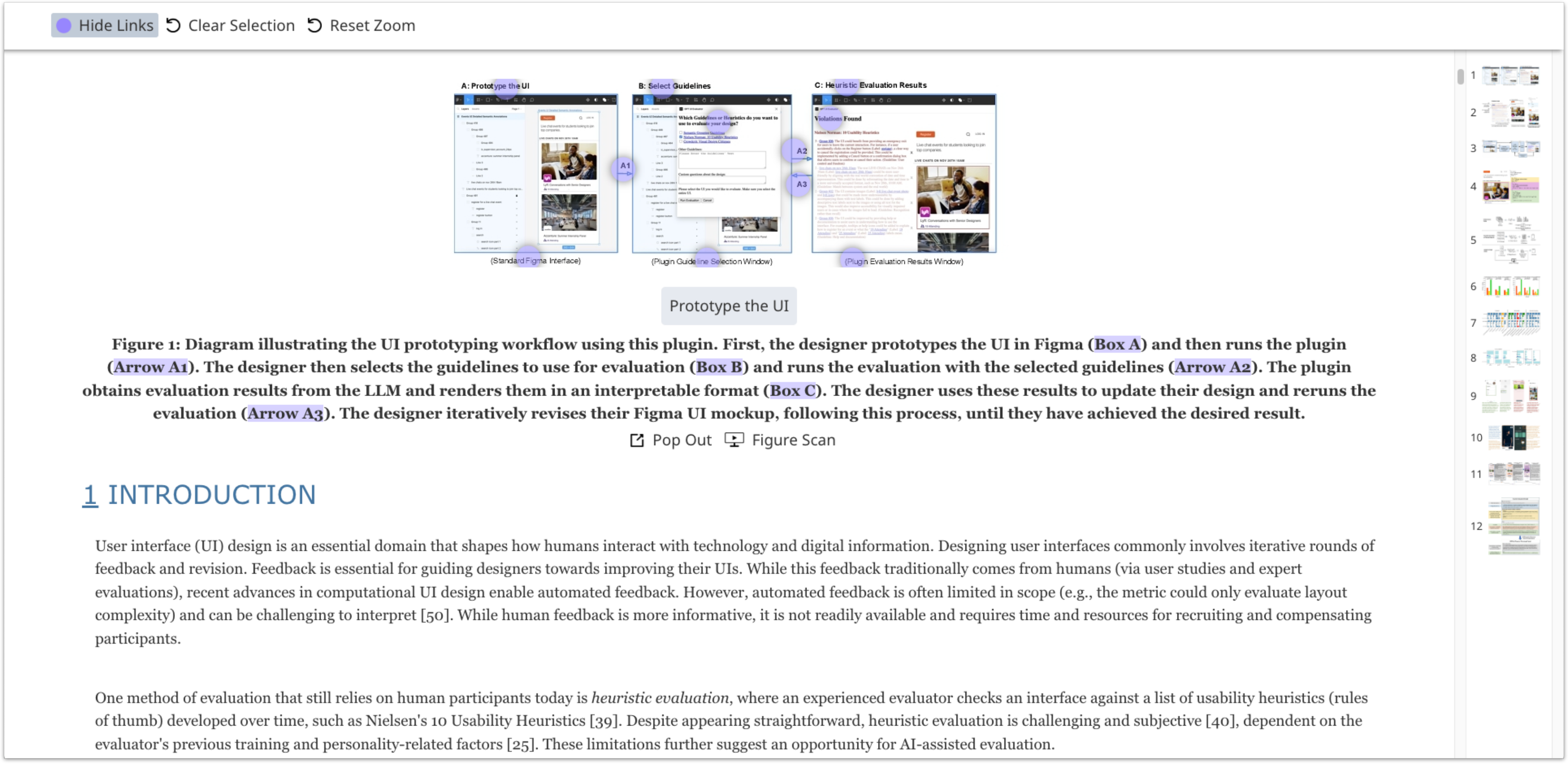}
    \caption{Opening view of interface. When it is initially launched, the interface shows a toolbar across the top and the visual index along the far right, with the paper taking up the majority of the screen. ``Linking mode,'' which displays the purple points and highlights, is activated by default. Main affordances are detailed in Section \ref{sec:system_affordances}.}
    \Description{Screenshot of interface with figure points and highlighted phrases.}
    \label{fig:ss1}
\end{figure}

Our interface integrates entities and links into research papers by augmenting HTML papers from the ACM Digital Library (see Figure \ref{fig:ss1}). In this section, we discuss the interface's affordances. The affordances are illustrated in the accompanying figures and \href{https://alyssahwang.com/projects/intfigs}{demo video}\footnote{https://alyssahwang.com/projects/intfigs}.

\smallskip

\begin{figure}[H]
    \centering
    \includegraphics[width=\linewidth]{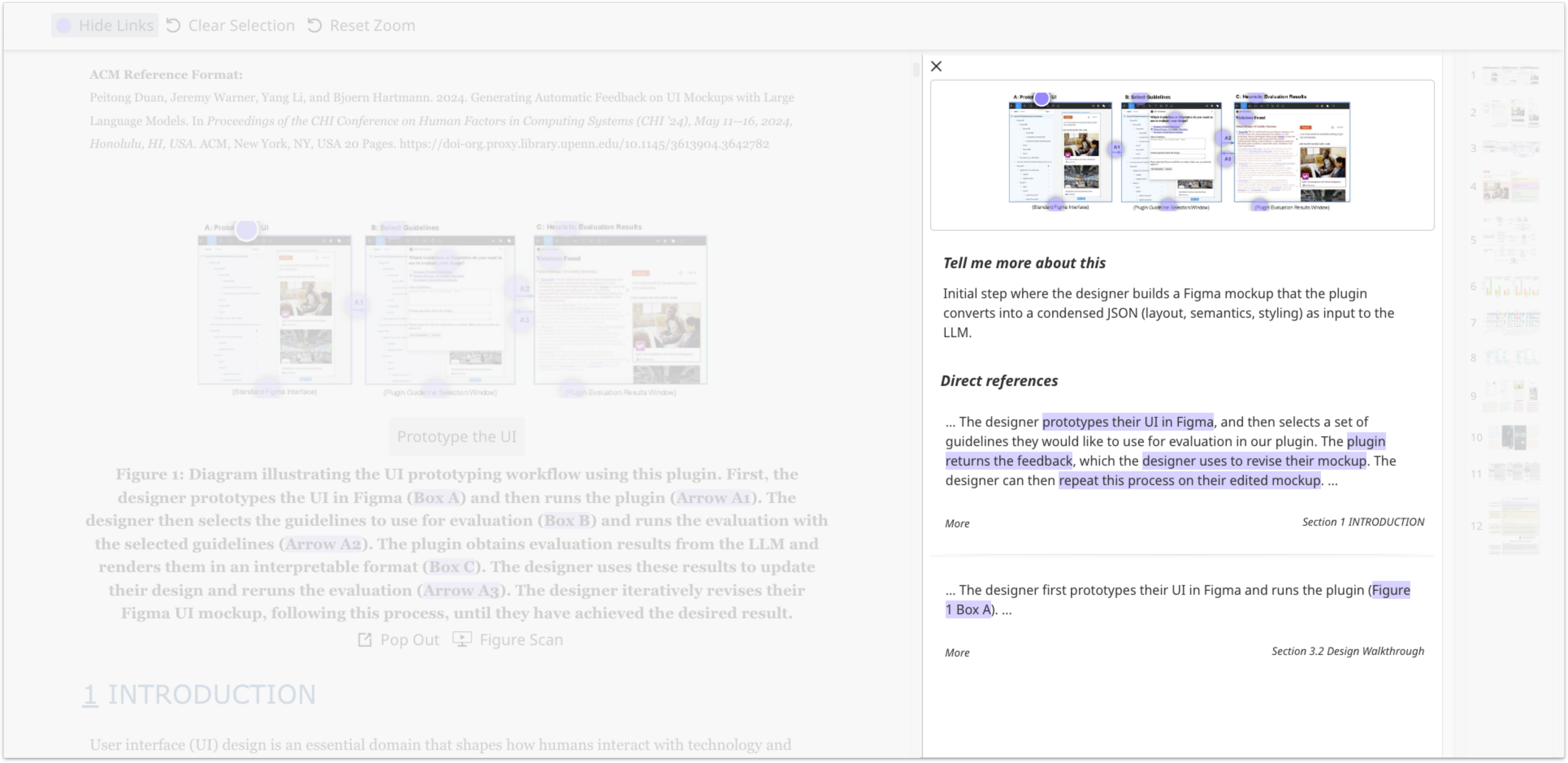}
    \Description{Screenshot of interface with spotlight on the reference panel.}
    \caption{Consolidation via reference panel. The reference panel offers a parallel working area for additional information. It opens to the right of the paper when the user clicks on a figure point or highlighted phrase. It is persistent, allowing users to continue scrolling in the main paper while providing quick access to a zoom- and pan-enabled copy of the figure (top), description of the selected entity (``Tell me more about this''), and links to related passages (bottom). The passages can be expanded and read in the reference panel without needing to scroll. When desired, the user can click on the passage to jump directly to it in the main paper.}
    \label{fig:rp2}
\end{figure}

\paragraph{Consolidation.} The figure points, highlighted phrases, and entity descriptions come together in a persistent \textbf{reference panel} (see Figure \ref{fig:rp2}). This panel appears to the right of the main paper when the user clicks on a figure point or highlighted phrase. It contains a functional copy of the corresponding figure, the description of the selected entity, and links to direct references (extracted during text processing) and other related passages (extracted during description generation). The links to direct references and other related passages can be expanded to display the entire paragraph inside the reference panel. The reference panel stays open until the user closes it, giving them easier access to information from throughout the paper.

\begin{figure}[htbp]
    \centering
    \includegraphics[width=\linewidth]{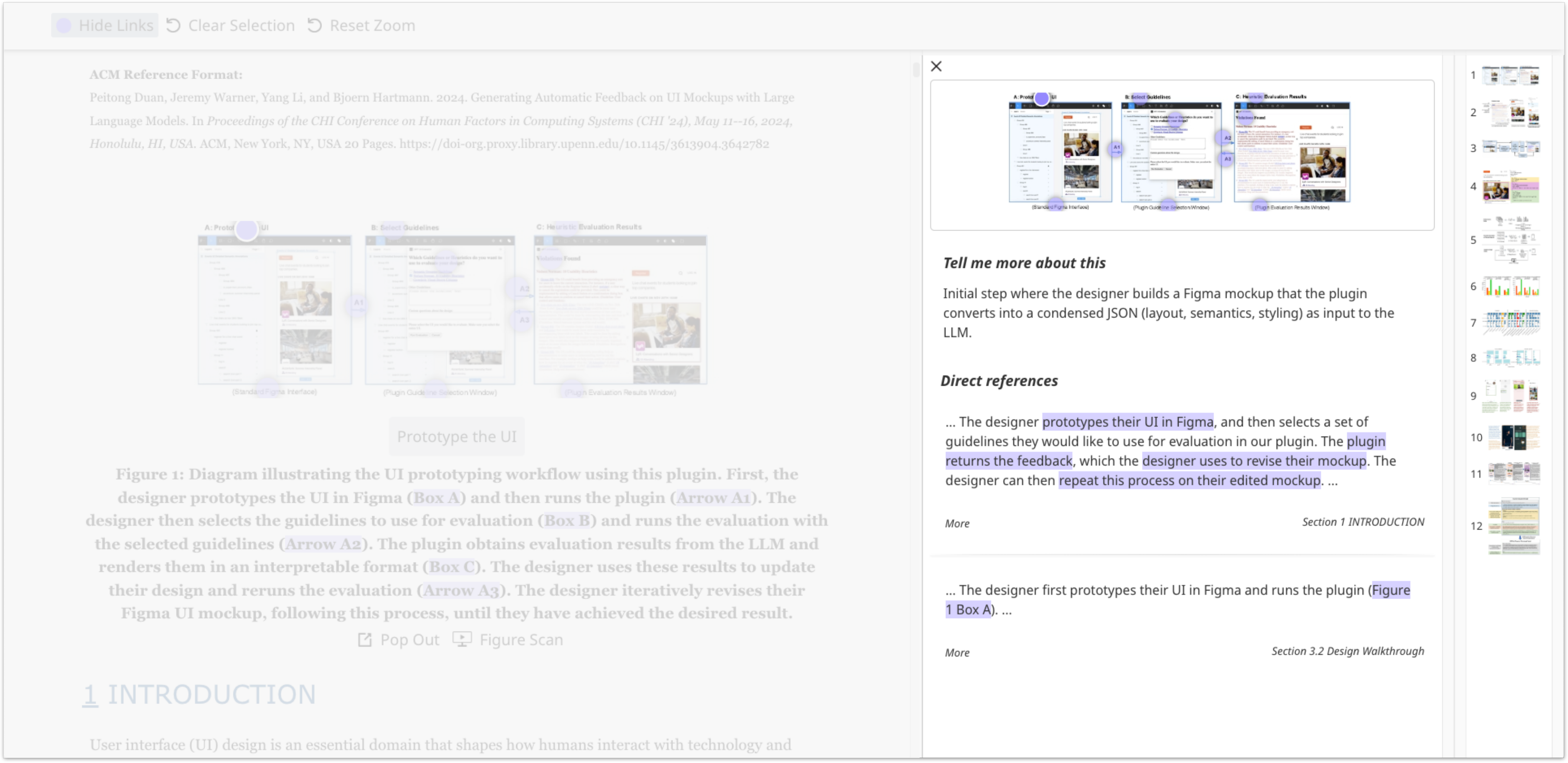}
    \Description{Screenshot of interface with spotlight on reference panel and visual index.}
    \caption{Navigation via passage links and visual index. Our interface offers two additional affordances for navigating papers, both of which involve directly jumping to a meaningful location. The first way is to click on a passage excerpt in the reference panel (see Figure \ref{fig:rp2} for more details on the reference panel). The other navigational affordance is the visual index on the far right. Users can click on one of the figures in the visual index to jump to it in the main paper.}
    \label{fig:nav}
\end{figure}

\FloatBarrier

\paragraph{Navigation.} The reference panel has a second purpose: helping users move directly to specific pieces of information. Clicking on a link to a direct reference or other related passage in the reference panel will automatically scroll the window to that location in the main paper. Another way our interface supports navigation is through the \textbf{visual index}, which displays small copies of figure images on the far right side of the interface. These images can be clicked to scroll directly to the figure in the paper, like a graphical alternative to a table of contents.

\begin{figure}[htbp]
    \centering
    \includegraphics[width=\linewidth]{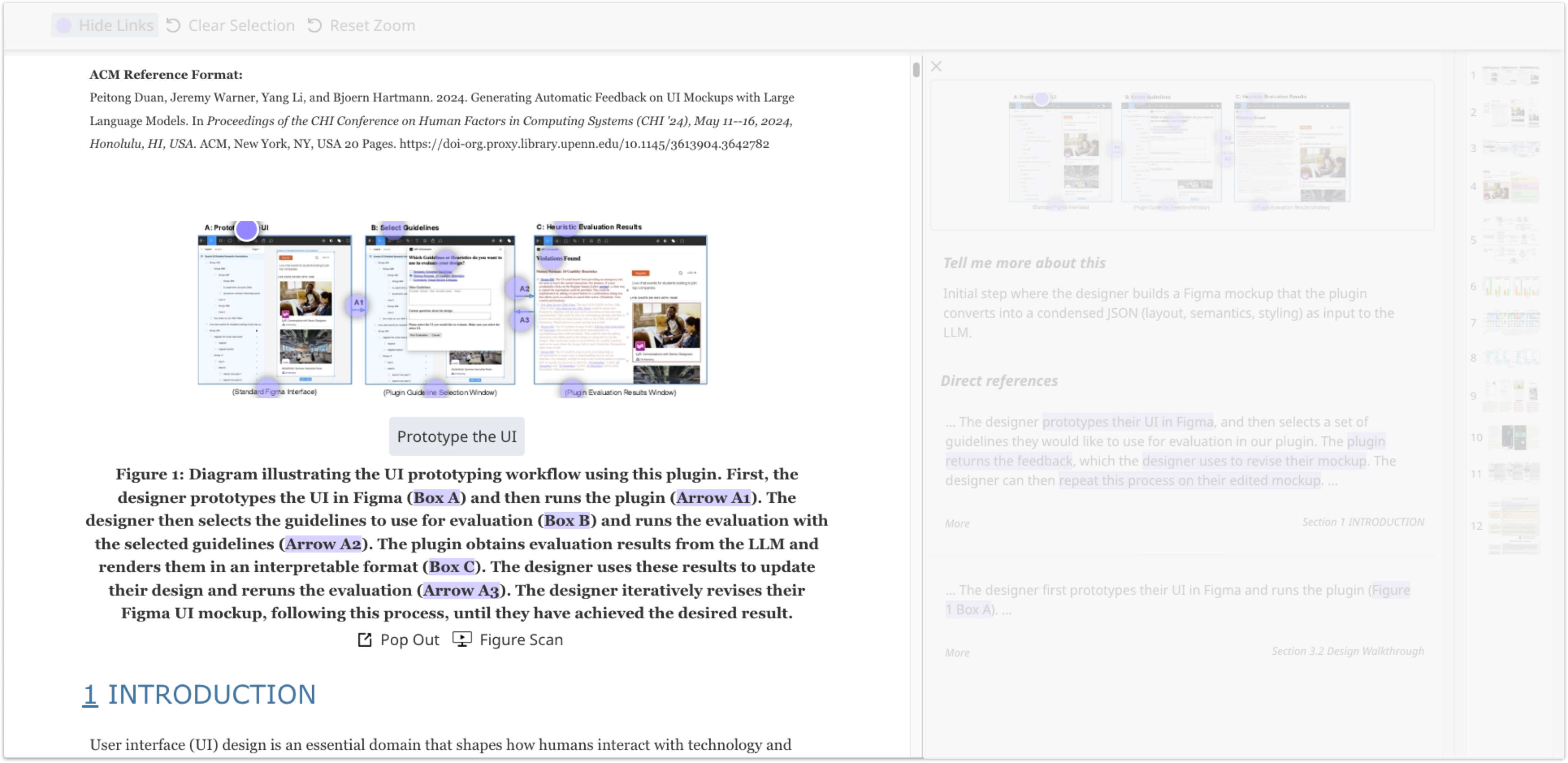}
    \Description{Screenshot of interface with a spotlight on the main paper.}
    \caption{Exploration via points and phrases. Interactive figure points and highlighted phrases allow users to explore visual and textual entities with their mouse. When linking mode is activated, each figure displays semi-transparent points that become opaque as the cursor approaches. Hovering over a figure point activates the linked phrase and vice versa to direct the user's attention. Clicking on a point or phrase opens and populates the reference panel with additional information (see Figure \ref{fig:rp2}).}
    \label{fig:figure}
\end{figure}

\FloatBarrier

\paragraph{Exploration.} The figure points and highlighted phrases are interactive, allowing users to hover over and click on them (see Figure \ref{fig:figure}). As previously discussed, clicking on a point or phrase opens the reference panel populated with other linked entities. Hovering over a highlighted phrase activates the linked point in the figure, which can be useful for quickly seeing where in the figure a particular caption entity is displayed. The links are bidirectional, so hovering over a highlighted phrase will make the linked points appear.

\begin{figure}[htbp]
    \centering
    \includegraphics[width=\linewidth]{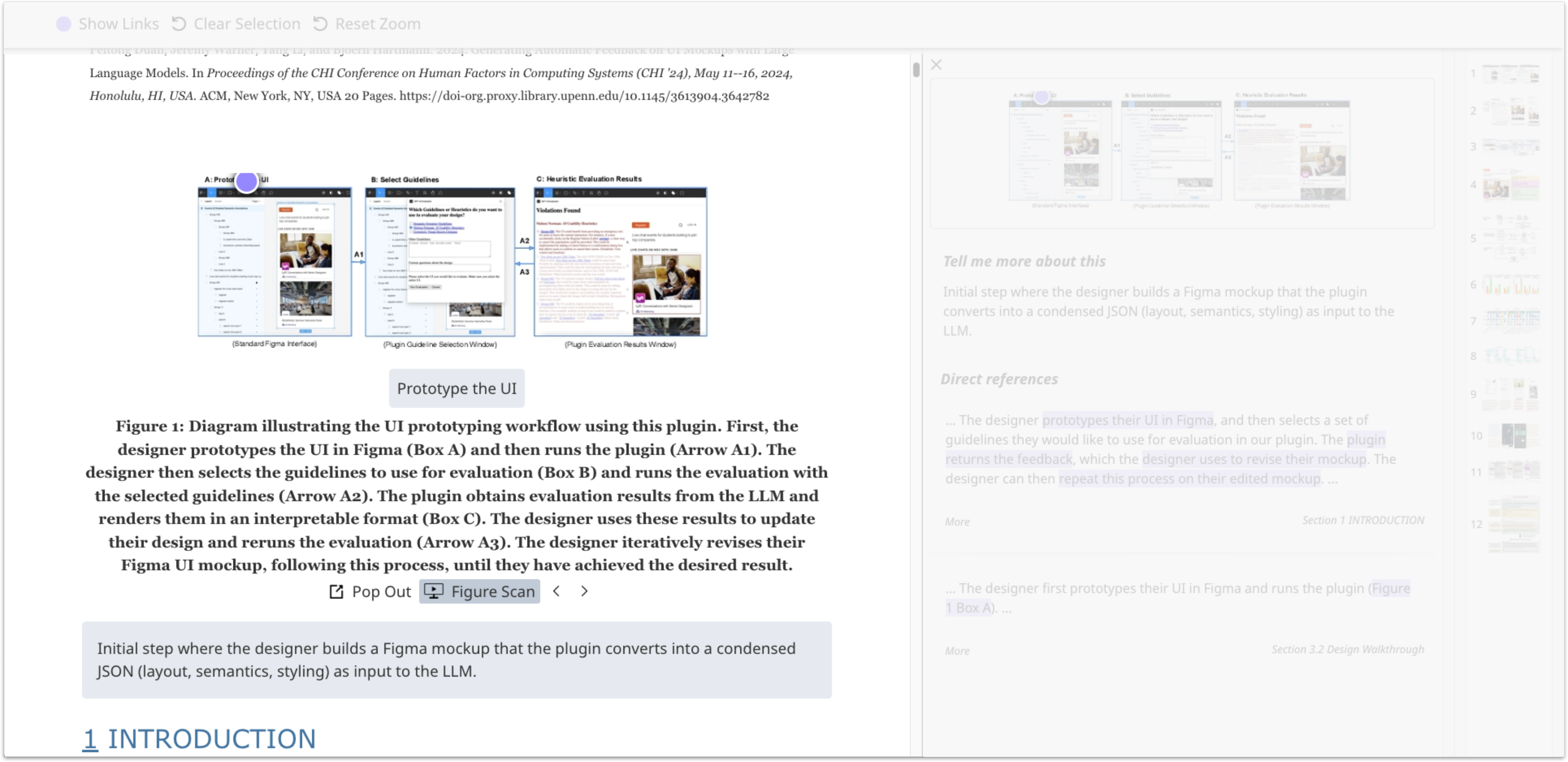}
    \Description{Screenshot of interface with spotlight on figure + entity description underneath caption.}
    \caption{Decomposition via figure scan. Figure scans provide more structured guidance by pointing at one entity at a time with a supplemental description beneath the caption.}
    \label{fig:scan}
\end{figure}

\FloatBarrier

\paragraph{Decomposition.} By default, the interface shows all figure points and highlighted phrases in a semi-transparent state. The default view naturally leans toward open-ended exploration, so we added \textbf{figure scans} for more structured guidance (see Figure \ref{fig:scan}). Activating the figure scan for a figure shows the points for an individual entity at a time (an entity can appear more than once in an image). The user can step backwards and forwards through the figure scan, which directs their attention to specific portions of the figure at each step. The entity description is reproduced under the figure caption as supplemental information, echoing a step-by-step tutorial. The reference panel also opens and automatically updates at each step, providing quick access to additional information.

\subsection{Insights from Iterative Design}
\label{sec:design_lessons}

Developing the framework and interface involved several cycles of iterative design, prototyping, and feedback. While most of the interface components were grounded in existing interaction paradigms, our goal was to adapt and integrate them to support information synthesis in long, multimedia documents. During this process, we identified several recurring challenges and design insights that shaped the final articulation of the framework.

\smallskip

\textit{Avoid visual overload.}
Figures containing too many interactive points created unnecessary visual clutter. Points should emphasize only key elements of a figure or be visually coded (for instance, by level of detail) to support scalable navigation without distraction. Refining the AI pipeline to focus on the most important entities that are visibly present in the figure and directly mentioned in the text improved usability.

\textit{Supplement, not replace.}
We experimented with different ways of presenting step-specific descriptions in earlier versions of the figure scan. Initially, the figure scan descriptions were intended to replace the original caption in a step-by-step guided walkthrough, but pilot participants expressed wanting to see the original text. We now present descriptions as supplementary to the original caption rather than replacing it.

\textit{Reframe based on technical capacity.}
Language model performance at the time prevented us from reliably producing the comprehensive, in-depth ``walkthroughs'' we had originally designed. We adapted walkthroughs into ``figure scans,'' which represent lightweight versions of walkthroughs that still provide a structured sequence of figure contents.

\textit{Consolidate contextual information.}
The reference panel evolved to unify multiple layers of information: the figure itself, labeled points, generated descriptions, direct references, and other related passages. Providing both precise (original) and abstracted (summarized) information within a single unobtrusive space aligned with the framework's goal of integrating distributed information.

\textit{Embed, rather than float, visual content.}
Earlier designs experimented with a floating pop-out figure, but users found it cumbersome to manage. Embedding the interactive figure within the reference panel provided stability, while zooming and panning supported fine-grained inspection.

\textit{Distinguish types of textual context.}
Pilot users requested clearer separation between explicit figure mentions and general context passages. Categorizing them as ``direct references'' and ``other related passages'' helped users understand what to expect from the different links in the reference panel.

\section{Design Validation Study}
\label{sec:thinkaloudstudy}

We assess the usability of our framework through a qualitative think-aloud study. The study focused on reading behavior, navigation strategies, and perceived integration of text and figures. All procedures were approved by the University of Pennsylvania Institutional Review Board.

\subsection{Participants}
We recruited 14 participants through academic and social media channels. To ensure that participants could meaningfully interpret complex research papers, we targeted researchers in human-computer interaction or related fields with prior experience in reading papers. Most participants (8/14) were PhD students, along with two undergraduates, two master's students, one research scientist, and one professor. Participants were compensated with a \$30 USD virtual gift card.

\subsection{Procedure}
Each study session lasted approximately one hour on Zoom. After obtaining informed consent, we demonstrated the use of our interface with an example paper \citep{hou_my_2024}. We introduced all major affordances derived from the framework: showing and hiding bidirectional links, interacting with figure points and highlighted phrases, stepping through figure scans, accessing the reference panel, and navigating with the visual index. Participants then completed short tutorial exercises on the same practice paper to familiarize themselves with the interface.

The main task was to read a different research paper~\citep{duan_generating_2024} while using our interface and thinking aloud. Participants were instructed to verbalize their thoughts, focusing on how the interface affected their understanding of figures, the connections between text and visuals, and their overall sense of document structure. They were encouraged to work naturally, reading as they typically would, while reflecting on the experience in real time.

Participants read for approximately 25 minutes and then elaborated on their experiences in a semi-structured debrief interview lasting about 20 minutes. The interview explored participants' perceived utility of the interface features, preferred interaction patterns, and moments of confusion or insight. Throughout the sessions, we collected screen and audio recordings, interaction logs, and questionnaire responses. 

\subsection{Observations}
In this section, we discuss our observations of participants interacting with framework affordances. Similar to the formative study, we analyzed our notes and transcripts following the established thematic analysis method \citep[Chapter 5]{blandford_qualitative_2016}, with themes validated by an external evaluator. We refer to participants with pseudonyms T3--T16; T1 and T2 were pilot participants who were excluded from main analysis.

\smallskip

\textit{Engagement with figure points.}
Figure points played a noticeable role during the reading sessions. Several participants (T8, T9, T12, T13) described the points as helpful for maintaining attention and retrieving contextual details. T13 noted that the feature ``made me have a better attention span,'' while T8 remarked, ``I truly love clicking on these points and seeing more about what it’s about.'' Some participants, however, found figures with too many points visually overwhelming, reinforcing the design lesson on managing visual density.

\textit{Information integration.}
The reference panel was another commonly used part of the interface. Most participants (8/14) used it to display figures and text side-by-side for cross-referencing. T8 described it as ``most useful for reminders\ldots{} to juxtapose the text with figures that were relevant and find related passages.'' Another participant reported that it was helpful because ``reading horizontally is more fluid than reading vertically'' (T12). Based on the participants' reactions, the reference panel seemed to support more seamless shifts between visual and textual details.

\textit{Elaboration.}
Half of the participants accessed the AI-generated descriptions of figure entities. For some participants, the descriptions filled in missing explanations or summarized visual content. T4, for example, appreciated that the descriptions ``explain what I should be taking away,'' while T13 described them as ``much friendlier'' and easier to process than the original content. Others, however, perceived them as overly general or repetitive (T3, T6, T8, T11, T15). Participants sometimes used the passage links in the reference panel to verify the descriptions, illustrating a new use for the AI-driven augmentations that we had not expected during the design process.

\textit{Navigation and search.}
The visual index also offered opportunities for information search. Originally just a simple navigation feature, the visual index helped some readers develop an understanding of the paper's overall structure. Participants such as T3 and T14 used it to gain a ``bird's-eye view'' of the paper and to locate figures more efficiently. For T8, it supported targeted exploration: when reading a statistic mentioned in one figure, she immediately jumped to a histogram to gather more context. She reported that the thumbnails in the visual index allowed her to search for follow-up information more efficiently than usual. During these sessions, the visual index showed potential to support search and navigation for non-linear, inquiry-driven reading.

\textit{Progressive visualization.}
The figure scan feature was used by ten participants in a few different ways. T15 treated it as a guided introduction before reading the figure on his own, while T11 used it to check his understanding afterward. T3 also reported that the scan ``gave me a mental model of how you're supposed to read the figure.'' The most commonly mentioned strength of the figure scan was the way it ``hides all the complexities and really just has you focus on one thing'' (T14). T15, T7, and T6 also expressed appreciation for minimizing the number of details in figure scans. Figure scans represented a structured alternative to the default view of figure points, both of which supported participants in different ways during our user study.

\section{Comparative Study}
\label{sec:eval}

To complement the qualitative findings from the think-aloud study (see Section \ref{sec:thinkaloudstudy}), we conducted a controlled between-subjects study to investigate the effects of our framework on reading outcomes. We measured performance on a reading quiz, time to completion, and perceived cognitive load for participants who read the same research paper using our interface or a baseline. We recruited senior undergraduates in engineering to gain a better sense of generalizability. This study was exempt by the University of Pennsylvania Institutional Review Board.

\subsection{Participants}

We recruited a total of 56 students from the University of Pennsylvania to participate in our study. All participants were recruited from the Computer and Information Science senior design course through a post on the class discussion forum. They were offered three points of extra credit toward their final grade as compensation. 

Of the 56 participants, 38 participated in pilot studies during our extensive study design process. The pilot participants helped us adjust study procedures, experiment with several versions of the reading quiz, and resolve technical issues before our official sessions. The 38 pilot participants were not included in the following demographic information.

We collected final data from the remaining 18 participants, who were academically similar to each other: 13 were undergraduate seniors while 5 were accelerated master's students. All 18 participants had a background in engineering, majoring or double-majoring in computer and information science (14), math (2), finance (1), or systems engineering (1) in particular. In addition, 4 participants were enrolled in Penn's dual-degree Jerome Fisher Program in Management and Technology (M\&T).

Participants had advanced standing in their academic programs but were relatively new to research. They reported a mean of 1.15 years of research experience ($\sigma = 1.03$) and a mean of 3.94 out of 7 ($\sigma = 1.39$) for familiarity with the research field of the stimulus paper. The participants seemed slightly more comfortable with reading a computer science research paper in general, reporting a mean of 4.61 out of 7 ($\sigma = 0.78$). The participants' academic and research characteristics indicated that they had enough background knowledge to read the stimulus paper while likely being open to additional reading support.

Beyond academics and research, participants were balanced by gender (9 female, 9 male) and ranged from 19 to 21 years old ($\mu = 21.06$, $\sigma = 0.64$). All participants reported native English reading proficiency except for two, who described themselves as intermediate readers.

\subsection{Study Probes}
\label{sec:eval_probe}

\newcommand{\parevaldesc}{Improving AI-generated descriptions.}

The baseline and experimental research probes were implemented as extensions of the prototype presented in Section~\ref{sec:design}. To create a consistent reading environment, all hyperlinks, including internal links to figures, tables, and references, were disabled. We also removed the default toolbars so that only the paper content remained. We regenerated all AI content using  GPT-5 with \texttt{thinking = "high"} and Molmo for improved data quality (see \textbf{\hyperlink{par:eval_desc}{\parevaldesc}} below). Additional implementation details are provided in Sections \ref{app:design_implementation} and \ref{app:eval_prompts}.

The baseline interface shares the same visual layout as the experimental version, besides the excluded augmentations. Since the functionality of the built-in browser search function (\texttt{Control-F}/\texttt{Command-F}) overlapped with the passage links in the reference panel, we disabled search in both interfaces.

Based on observations and feedback from the think-aloud study, we refined the experimental interface to address key usability challenges while maintaining the underlying framework and functionality. The following revisions were made before the final evaluation:

\smallskip

\textit{Reducing visual clutter.}
To simplify the visual environment and reduce cognitive load, we decreased the number of interactive points in some figures. Points without direct references or related passages were removed, and figure entities and links were regenerated using GPT-5 to produce a more coherent and focused set of augmentations (see Section \ref{sec:entitieslinks} for more information on the AI pipeline, which is the same except for the updated language model).

\hypertarget{par:eval_desc}{\textit{\parevaldesc}}
Participants in the think-aloud study noted that several of the AI-generated descriptions in the ``Tell me more about this'' feature were too generic or repetitive. To improve their quality, we regenerated descriptions using GPT-5 with \texttt{thinking = "high"} and revised about 5 of 90+ descriptions for grammar and clarity. Since our work focuses on the effect of high-quality integration on reading outcomes, not on benchmarking the capabilities of contemporary language models, we believe these changes did not substantially affect our results.

\textit{Enhancing image interactions.}
Participants frequently expressed a desire for greater control over figures, such as the ability to move or enlarge them. While the original interface supported zooming and panning in the reference panel, this view was too small to be practical. The revised interface introduces a ``popout'' figure that can be repositioned and resized while preserving the same interactive points and behaviors as the embedded figure.

\textit{Integrating the figure scan.}
In the prior version, the figure scan was perceived as separate from the rest of the interface. Some users also overlooked the ability to access detailed reference information while using the scan. To create a more cohesive experience, the final version automatically opens the reference panel when a figure scan is activated. The contents of the panel dynamically update as readers step through the scan.

\subsection{Procedure}
\label{sec:eval_procedure}
Each session lasted approximately one hour over Zoom, with participants using Google Chrome to access their assigned reading interface. After reviewing the study purpose and obtaining consent for audio and screen recording, we provided a live demonstration of the baseline or experimental interface to which the participant was randomly assigned. The baseline tutorial instructed participants to read by scrolling through the paper, while the experimental tutorial introduced the figure points, highlighted phrases, reference panel, related passage links, figure scans, and popout figures. The tutorial paper for the comparative study was the same as the one for the think-aloud \citep{hou_my_2024}.

During the main task, participants were given 25 minutes to read the main activity paper \citep{duan_generating_2024} with their assigned interface and complete an open-book quiz about it. The quiz consisted of ten short-answer questions distributed across the paper's text and figures (see Section \ref{app:eval_questions} for full quiz content). Participants were told to answer each question as accurately and quickly as possible within the time limit, which was enforced by the facilitator. Afterward, the facilitator conducted a semi-structured interview and administered a questionnaire about participants' backgrounds and experiences (see Section \ref{app:eval_questions} for questionnaire items).

\subsection{Data Collection and Analysis}
\label{sec:eval_analysis}

We collected several forms of data during each user study session: responses to the short-answer quiz, time to completion as measured by Qualtrics, self-reported ratings on the NASA Task Load Index (NASA TLX), transcripts of interviews, and screen recordings. The transcripts and recordings were produced by Zoom. NASA TLX, demographic information, and preference ratings were collected through an anonymous Google Forms questionnaire at the end of each session.

\textit{Response quality.} We evaluated responses to the short-answer quiz for correctness and completeness, which we collectively define as response ``quality.'' Two external annotators independently scored each response according to a rubric, which can be found in Section \ref{app:eval_rubric}. Both annotators were PhD students with backgrounds in research and computer science at the University of Pennsylvania. Responses received 2 (error-free and fully detailed), 1 (error-free but missing details), or 0 (containing errors) points for a maximum total score of 20 points per quiz. In a few instances\footnote{1 baseline response from questions 9 and 10; 1 experimental response from questions 8, 9, and 10.}, responses were missing because participants ran out of time. These scores were removed from analysis.

\smallskip

\textit{Time to completion.} We measured time to completion with the ``page submit'' feature of Qualtrics, which tracks the number of seconds between the participant opening the question and submitting their answer. Each question was displayed on its own page with blank pages in between them to record completion time more precisely. Durations for questions that were not attempted because of time limits were removed from analysis.

\textit{Cognitive load.} Participants rated six dimensions on a 7-point Likert scale: mental demand, physical demand, time pressure, performance, effort, and frustration. These dimensions were defined by the NASA Task Load Index (NASA TLX) and delivered through Google Forms after the reading session and debrief interview. The full phrasing of the questions is listed in Section \ref{app:eval_tlx}.

\section{Findings}
\label{sec:findings}

\subsection{Inter-annotator Agreement}
\label{eval:iaa}

Our annotators achieved Krippendorff's $\alpha = 0.75$, indicating that the quiz scores are of an acceptable quality for the remainder of our analysis \citep{krippendorff_reliability_2004} (see Appendix Figure \ref{fig:eval_iaa}).

\subsection{Response Quality}
\label{sec:eval_quality}

\begin{figure}
    \centering
    \includegraphics[width=0.75\linewidth]{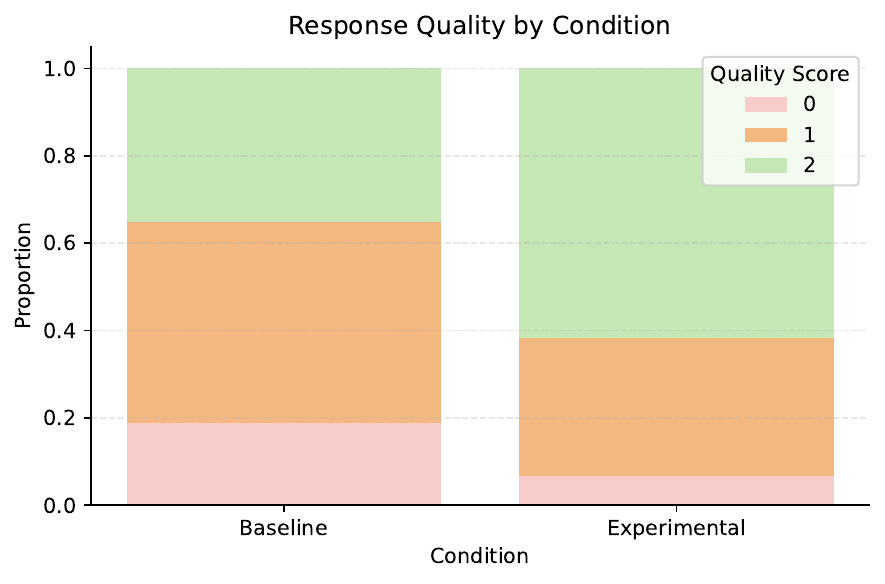}
    \caption{Proportions of scores for responses to the short-answer quiz. Each response received a score of 0 (incorrect), 1 (partially correct), or 2 (fully correct). The experimental condition shows an increased proportion of fully correct responses ($\chi^2(1) = 11.26$, $p = 0.00079$).}
    \label{fig:eval_accuracy}
\end{figure}

As judged by our external annotators, participants using the experimental interface submitted higher-quality answers than those using the baseline interface (see Figure \ref{fig:eval_accuracy}). The distributions of quality scores show a clear shift toward fully correct responses, demonstrated by the larger proportion of responses receiving full credit (``2'' points, green bar) and smaller proportion of responses containing errors (``0'' points, pink bar). These differences are statistically significant according to a Mann-Whitney U test ($U = 5042.5$, $p = 0.0000294$, $r = -0.33$).

Since ``quality'' in our case encompasses correctness and completeness, we can deduce that experimental participants submitted fewer errors and more detailed responses to the short-answer questions. The responses were also open-ended, meaning that participants needed to find and synthesize necessary details without additional writing support (i.e., from a chatbot). The statistically significant increase in response quality, therefore, suggests that our framework helped participants understand key parts of the paper. 

Improvements in response quality manifested differently across the ten quiz questions (see Appendix Figure \ref{fig:eval_perquestionacc}). Experimental bars are higher for nearly all questions, with tighter or comparable 95\% confidence intervals for several. This suggests that performance was often higher and more stable for participants who used our tool. Improvements in response quality were initially statistically significant for questions 6 ($U=75$, $p=0.00079$, $r=-0.852$), 8 ($U=52$, $p=0.0309$, $r=-0.444$), and 9 ($U=56$, $p=0.00272$, $r=-0.778$), with the results for questions 6 and 9 surviving Bonferroni correction (see Table \ref{tab:eval_questions}).

\begin{table}
\centering
\begin{tabular}{ccccc}
\toprule
Within caption & 2 paragraphs & 3 paragraphs & 4 paragraphs & Very far \\
\midrule
1, 2, 3 & 5, 9 & 4, 8 & 6 & 7, 10 \\
\bottomrule
\end{tabular}
\caption{Distance groups. Quiz questions were grouped by the distance in paragraphs between details required for a fully correct answer. Questions 7 and 10 represented distances of 62 and 7 paragraphs, respectively. The distance for question 7 was particularly far because it involved a figure in the appendix.}
\label{tab:eval_distancegroups}
\end{table}

\begin{figure}
    \centering
    \includegraphics[width=0.75\linewidth]{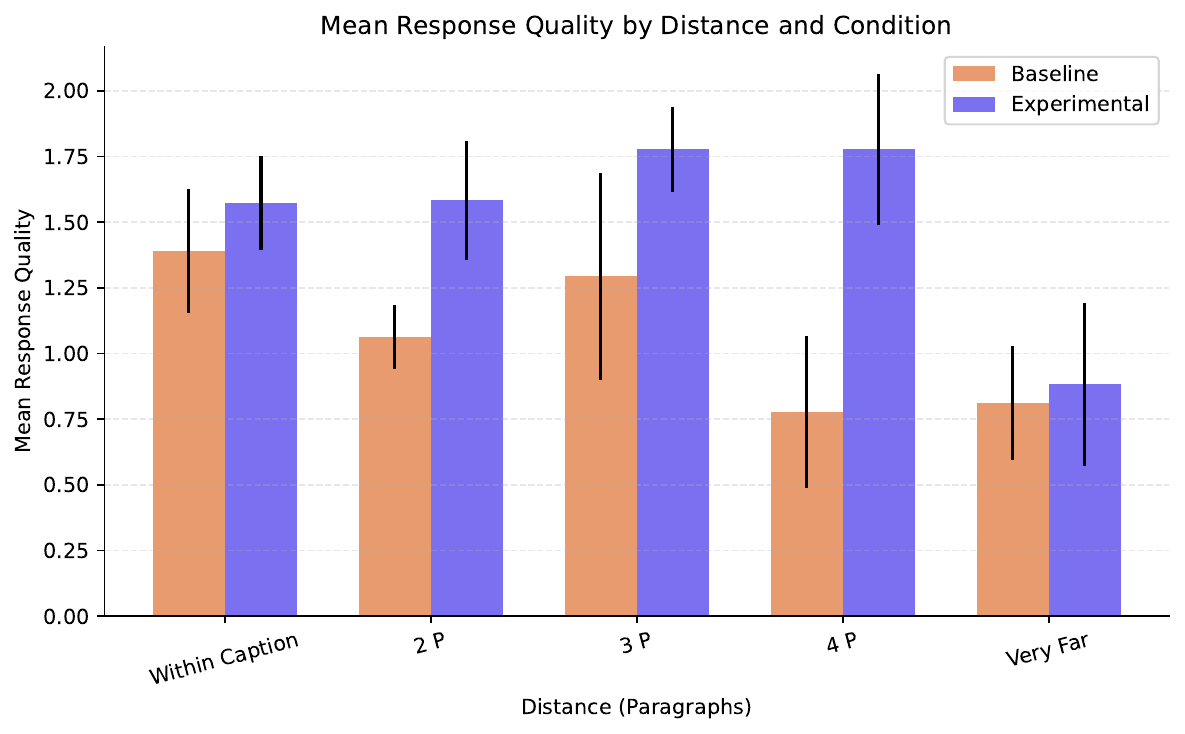}
    \caption{Response scores by distance group. Response quality was graded from 0 to 2 based on correctness and completeness. Questions were also grouped by distance in paragraphs between necessary details in the paper. We report mean response score per distance group with 95\% confidence intervals. Improvement in score was statistically significant for 2P and 4P after Bonferroni correction.}
    \label{fig:eval_distanceacc}
\end{figure}

A more structured understanding of response quality emerges when we group the questions by distance (see Table \ref{tab:eval_distancegroups} and Figure~\ref{fig:eval_distanceacc}). Each question required finding and reasoning about specific details related to a figure, so we quantified distance by counting the number of paragraphs between figures and corresponding details. Crucial details were four, four, and two paragraphs away from the relevant figures for questions 6, 8, and 9, respectively. Furthermore, improvement in response quality was initially higher for distances of two ($U = 223.5$, $p = 0.00115$, $r = -0.552$), three ($U=199$, $p=0.0471$, $r=-0.301$) and four ($U=75$, $p=0.00079$, $r-0.852$) paragraphs; distances of two and four paragraphs survive Bonferroni correction (see Table \ref{tab:eval_questions_distance}). These findings suggest that our framework may help more when necessary details are further apart.

The slight negative trend in question 7, however, suggests that the benefit of our framework may drop off at far distances. To answer question 7, participants usually started at Figure 4 in Section 3.3. They then needed to find an additional detail 62 paragraphs away, in an appendix figure. Participants may not have expected critical details to appear within other images or the appendix, leading to poor performance all around. Only one of the 18 participants received full credit from both annotators for this question. This participant was part of the experimental group, noticed the critical information in the ``Tell me more about this'' description in the reference panel, and immediately submitted the fully correct answer. Other participants scrolled the document or clicked through other augmentations before submitting an incorrect or incomplete answer. Our results show statistically significant improvement only when related details are two or four paragraphs away from each other, but additional studies may uncover other relationships between distance and performance.

\subsection{Time to Completion}
\label{sec:eval_time}

Despite improvements in response quality, our results showed no statistically significant evidence of difference in overall time to completion ($U = 3938.0$, $p = 0.320$, $r = -0.041$; see Appendix Figure \ref{fig:eval_time}). This finding is consistent when breaking the results down by question in Appendix Figure \ref{fig:eval_perquestiontime} and distance in Appendix Figure \ref{fig:eval_distancetime}.\footnote{Our analysis so far provides no evidence that our framework increases time to completion, which does not necessarily mean that time to completion with our framework is statistically equivalent to the baseline. To investigate further, we conduct TOST equivalence tests on mean time to completion with a 20-second equivalence margin. These tests did not demonstrate statistical equivalence overall, per question, or among distance groups (see Section \ref{app:eval_statsig}).}

We originally foresaw two possibilities for time to completion before beginning our studies. The first was that time to completion would increase for experimental participants because they spent extra time exploring augmentations. This could have led to increased frustration or mental demand as well. On the other hand, time to completion could have decreased because our framework sped up the process of searching for critical details. Our analysis eventually revealed no conclusive difference, possibly because experimental participants found relevant details more quickly but spent extra time double-checking other figure points or highlighted phrases.

\subsection{Cognitive Load}
\label{sec:eval_tlx}

Participants in the experimental setting needed to manage strictly more sensory input since our framework added a mean of 7 points per figure and a total of 116 highlighted phrases to the paper. However, like with time to completion, we observe no significant differences between the baseline and experimental conditions across all six NASA TLX dimensions (see Section~\ref{app:eval_tlx}). As shown in Appendix Figure~\ref{fig:eval_tlx}, participants in both groups reported similar levels of mental demand, physical demand, time pressure, perceived performance, effort, and frustration. None of these differences approached significance (all $p > 0.41$) and all effect sizes were small (all $|r| < 0.25$).\footnote{TOST equivalence tests revealed no equivalence in any of the TLX dimensions (see Table \ref{tab:app_eval_tlx_equiv}).}

\subsection{Preferences}
\label{sec:eval_preferences}

Overall, experimental participants reported being significantly more likely than baseline participants to choose their assigned interface for future reading tasks ($U = 59.0$, $p = 0.0498$). Among experimental features rated in Table \ref{tab:eval_preferences_features}, figure points and the reference panel were rated ``useful'' or ``very useful'' by all eight experimental participants who responded to this part of the questionnaire. Combined with the relatively high ratings for the highlighted caption and body phrases, affordances for visualizing intra-document links appeared to be the most well received. The popout and zoom/pan figures received the lowest ratings, suggesting that closely inspecting the images themselves was less helpful for this task. The response to figure scans and reference panel links was mixed, with some participants rating them positively and others not using them at all. These ratings align with what we learned in debrief interviews, during which participants emphasized success in searching for information driven primarily by the figure points.

\begin{table}[]
    \centering
    \begin{tabular}{lccccc}
    \toprule
        & Very useful & Useful & Somewhat useful & Not useful & Did not use \\
        \midrule
        Figure points & 6 & 2 & 0 & 0 & 0  \\
        Highlighted phrases in caption & 3 & 3 & 1 & 1 & 0 \\
        Highlighted phrases in body & 2 & 1 & 5 & 0 & 0 \\
        Figure scans & 2 & 2 & 1 & 0 & 3 \\
        Reference panel & 4 & 4 & 0 & 0 & 0 \\
        Links to passage in ref. panel & 0 & 5 & 1 & 0 & 2 \\
        Popout figure & 0 & 1 & 2 & 2 & 3 \\
        Zoom/pan figure & 0 & 0 & 1 & 4 & 3 \\
        \bottomrule
    \end{tabular}
    \caption{Ratings for experimental features. We report ratings for 8 of 9 experimental participants since one participant did not submit any. The figure points and reference panel were the most highly regarded features, while the popout and zoom/pan figures were among the least popular.}
    \label{tab:eval_preferences_features}
\end{table}

\section{Qualitative Insights}
\label{sec:eval_qual}

We present major qualitative insights based on debrief interviews and observations. The interview transcripts were analyzed following the thematic analysis \citep{blandford_qualitative_2016} and grounded theory methodologies. We assign participants anonymous identifiers starting with PB (for baseline) or PE (for experimental) followed by a number from 0 through 17. The baseline and experimental groups were balanced, with nine participants in each. Baseline participants consisted of PB0, PB5, PB7, PB8, PB9, PB11, PB13, PB14, and PB17, while experimental participants included PE1, PE2, PE3, PE4, PE6, PE10, PE12, PE15, and PE16. Some excerpts were lightly revised for clarity.

\smallskip

\textit{Baseline challenges.}
Baseline and experimental participants described their reading efforts differently. Five baseline participants (PB0, PB5,  PB7, PB11, PB14) mentioned time pressure as a central concern, in contrast to just one of the experimental participants (PE3). The ``decently long'' length of the paper contributed to ``a little bit of a time crunch,'' as PB11 noted. Besides time pressure, the concern mentioned by the most baseline participants (6/9) was ``finding the right information at the right place'' (PB13). Searching for information was particularly difficult when it was ``tucked away at the end of a paragraph'' (PB9) or ``present in a singular sentence in a singular section'' (PB11). This led to over half of baseline participants (5/9) to report a substantial amount of scrolling during their sessions (PB0, PB7, PB8, PB9, PB14). Searching for information was easiest when its location could be easily predicted, according to PB7, PB8, PB9, PB11, and PB14. As PB9 put it, ``if you know where to look, then you just look there.''

\textit{Experimental reactions.}
In contrast, experimental participants tended to frame their efforts in terms of using the interface, often with positive reactions. The majority of experimental participants (7 out of 9) identified figure points as the most helpful part of the interface. Figure points directly helped with the search for information, as PE2 described:

\begin{quote}
    Whenever a question referenced something like, for example, Figure 1 Box 2, I can easily go to Figure 1, Box 2, click the purple circle there, and then be able to get the direct information I need.
\end{quote}

\noindent Along with figure points, entity descriptions were among the most commonly praised parts of the interface (PE3, PE6, PE12, PE15, PE16). P15 liked the descriptions because they ``basically gave you the breakdown of what the figure was telling you'' while PE3 found them useful for ``putting the explanation of these figures into perspective, relative to the context of the paper.''

\textit{The (potential) power of a summary.}
Natural-language augmentations like entity descriptions could be a powerful part of future interfaces, especially since  summaries were the most requested feature across conditions when participants were asked to envision the perfect tool for understanding papers (10/18). Summaries could be used to support ``incremental understanding'' (PB5), ``[point] out relevant info'' (PE6), or identify ``important findings\ldots{} important methodology parts and important results'' (PB17). Some participants even wanted to interact directly with a language model to receive tailored explanations or references (PB0, PB5, PE6, PB11, PB13, PE15).

\textit{Opportunities for development.}
Experimental participants identified a few other areas for potential development. Caption phrases (3/9) and direct references (2/9) in particular were described as less useful than other parts of the interface. PE16 explained, ``I didn't need to click on the [linked caption phrases]\ldots{} I think the purple dots in the actual figure that link to the description on the side covered everything I needed to know.'' Direct references often linked to figure captions, compounding the redundancy. The quiz questions were mostly figure-centric, so the linked phrases could have been perceived as less useful because they needed less use. The popout and zoom/pan figures and figure scans were similarly considered less useful or used less often (see Table \ref{tab:eval_preferences_features}). Figure scans in particular seem like a promising direction for future work. Our figure scans were received well by experimental participants, but we gather more insights about them in longer or more high-stakes (e.g., preparing for a paper presentation) reading sessions.

\section{Conclusion}
\label{sec:conclusion}

In this paper, we developed and instantiated a framework  that supports the consumption of complex documents. Abstractly, this framework consists of entities (e.g., important concepts, claims, data) and links between them (e.g., explaining, providing an example, defining). The framework was instantiated as an augmented reading interface for research papers, which we evaluated through user studies. Our framework led to statistically significant improvements in reading quiz performance when compared to a baseline without evidence of increased time to completion or cognitive load.  Connections between details may be more helpful when they are further apart, as demonstrated by statistically significant improvement particularly for questions with key details two or four paragraphs away. More broadly, our work highlights the potential of treating complex documents as networks of entities and links, exposing opportunities for integrating information at fine granularities across document modalities.

\subsection*{Limitations}
Most of the development and evaluation of our framework was grounded in computer science research papers, leading to several limitations. The instantiation of our framework was implemented as an extension of the ACM (Association for Computing Machinery) Digital Library, which provides HTML versions of recent publications. Example papers we used during internal testing were mostly from human-computer interaction (HCI) or natural language processing research, further limiting the types of textual and graphical content we observed.

Our efforts to conduct a controlled study in Section \ref{sec:eval} introduced limitations as well. We used one HCI research paper \citep{duan_generating_2024} for all participants. The participants were similar to each other in terms of academic standing (upper undergraduate), research experience (median of one year), and English reading proficiency (all native except for two intermediate). We did recruit HCI researchers for the formative and think-aloud studies in Sections \ref{sec:formativestudy} and \ref{sec:thinkaloudstudy}, but expanding the range of users even more would be a worthwhile direction for future work. Sample sizes were small, limiting statistical power (especially for time to completion and cognitive load). The study involved a 25-minute reading session and open-book quiz, which may not reflect long-term or real-world reading behavior.

Although this work focuses on the interaction between users and our framework, generative AI was a substantial part of the process. The instantiation of our framework was enabled through GPT-4o (for earlier work in Section \ref{sec:design}), GPT-5 (for the final interface in Section \ref{sec:eval}), and Molmo (for both). We chose these models because they were easily accessible and performed well for our purposes, but they constrained our understanding of how well other currently available models might support our framework. This means that results may vary if the instantiation of our framework is populated with different models. 

\section*{AI Use Disclosure}
ChatGPT was used to help with outlining and revising this paper. Beyond writing, ChatGPT assisted with choosing statistical tests, drafting starter code for statistical analysis, and extracting insights from interview transcripts. Copilot (via Visual Studio Code inline tab-completion suggestions) and ChatGPT provided some coding and debugging support for the implementation of the interface; no ``vibe coding'' was used in this work. All LLM-generated content was reviewed and revised by an author before being included.

\begin{acks}
We would like to thank the participants of our formative, think-aloud, and comparative studies for their input on our work. We are especially grateful for assistance from Mari Thach, who handled logistics, and Liam Dugan, for his never-ending support.
\end{acks}

\bibliographystyle{ACM-Reference-Format}
\bibliography{references,references2}

\appendix

\section{Framework Implementation Details}
\label{app:design_implementation}

This appendix provides a detailed description of the technical pipeline used to implement the fine-grained augmentation prototype introduced in Section~\ref{sec:design_implementation}. The implementation consists of five stages: document preparation, entity identification and localization, text–visual linking, description generation, and interface augmentation.

\subsection{Document Preparation}
Each research paper was processed from its HTML version obtained from the ACM Digital Library. Working with HTML enabled direct augmentation through JavaScript and CSS. Using Beautiful Soup 4, we extracted all figures, captions, and body passages for downstream analysis.

\subsection{Entity Identification and Localization}
For each figure, GPT-4o was used to identify conceptually significant entities such as steps, results, or system components. Prompts instructed the model to avoid surface-level details (e.g., text labels or decorative shapes). Entity coordinates were then obtained using Molmo~\citep{deitke_molmo_2024}, which returns bounding-box coordinates in response to textual queries.

\subsection{Text–Visual Linking}
GPT-4o was subsequently prompted to identify phrases in the text referring to each detected entity. Both figure captions and body passages that explicitly mentioned the figure were processed. Because references varied in phrasing (e.g., “Step 1,” “Source Design Step”), the model was guided to reconcile synonymous expressions. Early iterations generated excessive or irrelevant links, which were reduced by limiting input to passages explicitly referencing each figure.

\subsection{Description Generation and Related Passages}
GPT-4o also produced short, context-aware descriptions for each entity, using the full text of the paper to provide concise conceptual explanations. During this process, the model retrieved additional passages containing related context, which were categorized as “other related passages” in the interface. These descriptions and passages supported the reference panel and figure scan features described in Section~\ref{sec:design}.

\subsection{Interface Augmentation}
The annotated document was rendered as an interactive web interface. The HTML text was augmented offline with \texttt{<span>} elements marking linked phrases. Figures were wrapped in \texttt{<svg>} containers, and circular overlays were added at Molmo-predicted coordinates. To manage visual clutter, points remained semi-transparent and became fully visible when the cursor approached, implemented using the Proximity JavaScript library. All augmentations were preprocessed locally to enable fast loading without live model calls.

\section{Original Large Language Model Prompts}
\label{app:design_prompts}

We provide the prompts used to generate the entities, links, and descriptions in Section \ref{sec:design}. We used the latest models available to us at the time of development, which included GPT-4o for the version of the interface used in the think-aloud study (Section \ref{sec:thinkaloudstudy}). These queries were made through the OpenAI API in the last quarter of 2024.

\subsection{Visual Entity Identification}
\label{app:design_prompts_image}

We used the following prompt to identify salient entities in figure images. We provided figures without additional context like captions.

\begin{lstlisting}[language=Python]
"""Help me label all of the items in this image. List all of the items one at a time with no additional context."""
\end{lstlisting}

\subsection{Textual Entity and Links Identification}
\label{app:design_prompts_text}

We used GPT-4o for the think-aloud study in Section \ref{sec:thinkaloudstudy} and regenerated content with GPT-5 for the research probes in Section \ref{sec:eval_probe}. We queried the OpenAI API with the figure caption or a body passage directly referencing a figure. We replaced \texttt{<entities>} with the list of visually identified entities and \texttt{<caption>} with the caption or referring passage.

\begin{lstlisting}[language=Python]
"""
You are a helpful data scientist.

I will give you the caption for a figure in a computer science research paper.

First, determine if the caption is describing multiple subfigures. If it is, extract the label of each subfigure, without unnecessary punctuation.

Then, determine if the figure is some kind of data visualization. If it is, follow these steps for each sentence:
1. Identify any labels related to the data visualization. Provide textual information as well, not just numbers. Include labels that are critical for someone to understand the figure without seeing it, like axis titles and data categories.
2. Skip labels that are related to data visualization in general.
3. Extract each label exactly, without parentheses or quotation marks.
4. Determine one of the entities below that this label represents. If none are appropriate, make a new one.
5. Move on to the ** RESPONSE FORMAT ** instructions.

If the figure is not some kind of data visualization, follow these steps for each sentence:
1. Check if the sentence contains labeled references to parts of the figure.
2. If the sentence contains labeled references, extract them exactly. If the label represents a range of items, produce each item in the range, even if you have to infer.
3. If the sentence contains no labeled references, identify the entities in it.
4. For each entity, try to guess if it refers to a particular part of the figure. Valid entities may be special names or possible quotations from the image text.
5. If you think the entity is referring to a particular part of the figure, extract it exactly.
6. Reread the caption again before providing your output. Add any references you might have missed.
7. Determine one of the entities below that this label represents. If none are appropriate, make a new one.
8. Move on to the ** RESPONSE FORMAT ** instructions.

** RESPONSE FORMAT **
Provide your response in valid JSON without any additional messages or information. The key should be the sentence and the value should be a list of pairs of extracted labels and entities. The label should be first and the entity should be second. Reproduce all text in the caption including the figure title in order in the JSON, using empty lists for sentences without references. If I concatenate all keys in the JSON, I should be the identical caption.
** LIST OF ENTITIES **
<entities>

** CAPTION **:
<caption>
"""
\end{lstlisting}

In the final sentence of the response format, ``I should be the identical caption'' is a typo for ``it should be the identical caption.'' We did not discover this typo until after data generation had been completed.

\subsection{Coordinates}
\label{app:design_prompts_coordinates}

We used a local version of Molmo-7b to identify coordinates of visual entities. We queried one entity at a time, replacing \texttt{\{target\}} in the prompt.

\begin{lstlisting}[language=Python]
"""Point at {target}."""
\end{lstlisting}

\subsection{Walkthroughs}
\label{app:design_prompts_walkthroughs}

In early versions of the interface in Section \ref{sec:design}, we used GPT-4o to generate step-by-step walkthrough tutorials of each figure. We used a two-part process: we first generated the text of the walkthrough and then identified links between entities and phrases within the walkthrough text. We replaced content in \texttt{<angle brackets>} with the corresponding value. These walkthroughs were generated through queries to the OpenAI API. In this early phase of development, we relied more on text-text matching since text-image matching performance did not fulfill our requirements at the time.

\begin{lstlisting}[language=Python]
"""
You are experienced in understanding figures in research papers.

**Task**: Generate a detailed step-by-step walkthrough of a figure in a research paper.

**Original Caption**: Here is the original caption of the figure from the paper, you can use it as the context for generating the walkthrough.
<caption>

**Context**: Here is the context for the figure from the paper. You can use it to generate the walkthrough.
<context>

**Requirements**:
1. Review the figure and generate a detailed, conceptual explanation of it.
2. Instead of repeating low-level visual details, talk about the high-level concepts and important ideas that the figure represents.
3. The explanations should be presented in a series of steps, where each step is a concise yet informative summary of some part of the figure.
4. The steps can be a few sentences in length. They should not be overly verbose, but they should not be too short or shallow, either.
5. You can incorporate details from the provided caption and context to help readers who have not read the paper understand the figure from the walkthrough.
6. The walkthrough should provide enough information for a reader to understand the figure out of context from the paper.
7. Provide each step of the walkthrough on a separate line without bullet points.
8. Provide the walkthrough without any additional context, explanation, introduction, or narrative. List the steps one line at a time with no bullet points.
9. Do not use bullet points of any kind. Give just the text of the walkthrough, one line at a time.
"""
\end{lstlisting}

\begin{lstlisting}[language=Python]
"""
You are good at explaining papers to people by pointing to the relevant parts of the figure.

**Task**:
Input: a description of a figure in a paper, its caption, and a "walkthrough" description of the figure.
Output: instructions for pointing to the parts of the figure relevant to each step in the walkthrough.

**Figure Information**:
Instead of showing the visual figure, I will provide you with textual information about the figure.

Figure number:
<figure_number>

Caption from the paper:
<original_caption>

Generated walkthrough:
<walkthrough>

**Requirements**:
1. For each step of the walkthrough, generate a list of element names that the step is referring to. Each element name is a phrase that describes a part or group of parts in the figure.
2. Be consistent with element names. Reuse element names you have already generated when possible.
3. Ensure those elements are clearly visible in the figure with appropriate granularity.
4. You should also extract the exact phrase without modifications from the sentence that refers to the element.
5. Phrases should not be nested in each other, i.e., a phrase should not be a part of another phrase.
6. When reproducing the walkthrough step for the JSON format, use the exact same step from the input without any modifications.

**Output Format**: provide your answer in JSON format with the following structure:
{
    "step_1": [["element1", "phrase for element1"], ["element2", "phrase for element2"], ...],
    "step_2": [["element3", "phrase for element3"], ["element4", "phrase for element4"], ...],
    "step_3": [],
    ...
}

Here is an example from a different paper. This is the example caption:
Figure 1: System architecture of CareCall, describing Ⓐ a chatbot conversing with users and Ⓑ a dashboard for teleoperators.

Here is the walkthrough:
The CareCall User initiates a call dialog.
The dialog is processed by the CareCall Chatbot.
The chatbot receives input and selects 20 relevant samples. These samples are drawn from an example dialog corpus.
The chatbot uses a pretrained LLM (HyperCLOVA). This model is trained on a large-scale Korean corpus, including blogs, social media, and forums.
The CareCall Dashboard receives the complete dialog.
The dashboard includes a User State Classifier. It provides summarized states. These states include health metrics and emergency alerts. A teleoperator can intervene on demand.

Here is acceptable output for this example figure:
{
    "The CareCall User initiates a call dialog.": [["user", "CareCall User"]],
    "The dialog is processed by the CareCall Chatbot.": [["CareCall Chatbot", "CareCall Chatbot"]],
    "The chatbot receives input and selects 20 relevant samples. These samples are drawn from an example dialog corpus.": [["CareCall Chatbot", "chatbot"]],
    "The chatbot uses a pretrained LLM (HyperCLOVA). This model is trained on a large-scale Korean corpus, including blogs, social media, and forums.": [["LLM", "pretrained LLM (HyperCLOVA)"]],
    "The CareCall Dashboard receives the complete dialog.": [["CareCall dashboard", "CareCall Dashboard"]],
    "The dashboard includes a User State Classifier. It provides summarized states. These states include health metrics and emergency alerts. A teleoperator can intervene on demand.": [["dashboard", "CareCall dashboard"], ["User State Classifier", "User State Classifier"], ["teleoperator", "teleoperator"]]
}

Use the instructions and example above to generate your response for the given figure. Remember to produce valid JSON for phrases that directly and specifically refer to parts of the figure.
"""
\end{lstlisting}

\section{New Large Language Model Prompts}
\label{app:eval_prompts}

GPT-5 was released soon before our final evaluation in Section \ref{sec:eval}. We used it through the OpenAI Playground to regenerate visual entities and walkthroughs/figure scans. Link generation and coordinates identification were the same as the original process, except for replacing GPT-4o with GPT-5 for links.

\subsection{Visual Entity Identification}

We used the following prompt in a developer message provided the PDF of the full paper as an attachment in a user message.

\begin{lstlisting}[language=Python]
"""For each figure, identify important elements within the image. Focus on elements with conceptual significance, not visual details. Label them descriptively, not just by repeating the text. For graphs, identify elements that have important trends, not just individual points. The labels should identify objects, not describe them. Provide your response in JSON format with "fig#" as the key and a list of elements (strings) as the value. Replace # with the figure number."""
\end{lstlisting}

\subsection{Figure Scans}

Originally framed as walkthroughs, ``figure scans'' were provided by generating descriptions of the visual entities instead of generating entirely new content and entities. This prompt was provided in a developer message.

\begin{lstlisting}[language=Python]
"""You are a senior PhD student preparing to present a paper at reading group. You assume that no one else has read the paper that deeply, so your answers will be concise, include critical context, and focus on conceptual understanding. Do not talk about visual or high-level characteristics of elements in a figure. In one sentence, describe the listed entities in the figure. Provide additional details from far away in the paper. Provide your response in JSON format with the element name as the key and the description of the value."""
\end{lstlisting}

\section{Questionnaire}
\label{app:eval_questionnaire}

Participants answered the following questions through an anonymous Google Form at the end of their sessions. The starred (*) questions provided an open text response for participants to self-describe. All other questions accepted one choice from a list of prepared answers (listed in parentheses).

\begin{itemize}
  \setlength{\parskip}{0pt}
    \item How useful were the following features for this task? (Did not use, Not useful, Somewhat, Useful, Very useful)
        \begin{itemize}
            \item Points
            \item Caption phrases
            \item Body phrases
            \item Figure scans
            \item Ref panel (general)
            \item Passages in ref panel
            \item Popout figure
            \item Zoom/pan figure
        \end{itemize}
    \item How likely would you be to choose this interface for reading research papers? ($1 =$ Not at all likely to $7 =$ Very likely)
    \item Year at Penn*
    \item Major(s)*
    \item Minor(s)*
    \item Age in years*
    \item Gender*
    \item How would you describe your English reading proficiency? (Basic, Intermediate, Advanced, Other)
    \item How many years of research experience do you have in computer science?*
    \item How would you describe your typical level of comfort when reading a CS research paper? ($1=$ Very uncomfortable to $7=$ Very comfortable)
    \item How familiar were you with the field of the paper you read today? ($1=$ Very unfamiliar to $7=$ Very familiar)
\end{itemize}

\section{NASA Task Load Index}
\label{app:eval_tlx}

We used the NASA Task Load Index (NASA TLX) in Section \ref{sec:eval_procedure} to measure cognitive load. We replicated the phrasing of the original TLX and recorded ratings from 1 (very low) to 7 (very high) (except for item \ref{list:tlx_performance}, from perfect to failure). Our abbreviations are indicated in parentheses.

\begin{enumerate}
\setlength{\parskip}{0pt}
    \item (mental demand) How mentally demanding was the task?
    \item (physical demand) How physically demanding was the task?
    \item (time pressure) How hurried or rushed was the pace of the task?
    \item \label{list:tlx_performance} (performance) How successful were you in accomplishing what you were asked to do?
        \begin{itemize}
            \item (our addition) Note that ``perfect'' is on the left at position ``1.''
        \end{itemize}
    \item (effort) How hard did you have to work to accomplish your level of performance?
    \item (frustration) How insecure, discouraged, irritated, stressed, and annoyed were you?
\end{enumerate}

\section{Short-Answer Quiz}
\label{app:eval_questions}

See Table \ref{tab:app_eval_questions} for the quiz questions described in Section \ref{sec:eval_procedure}.

\begin{table}[]
    \centering
    \begin{tabular}{cp{13cm}}
    \toprule
    Q & Text \\
    \midrule
        1 & What does ``Box A'' in Figure 1 represent? Describe what the user and system would be doing. \\
        2 & What does the Heuristic Evaluation Results Window allow users to do as shown in Figure 1? \\
        3 & In Figure 1, Arrow A3 forms a loop between Box B and Box C. Describe what makes this part of the process cyclical. \\
        4 & Figure 2 shows several ways in which feedback is displayed to the user in an interpretable rendering. Explain what happens when the user clicks on the ``X'' button of a violation. \\
        5 & Describe the ``Condensed JSON Representation'' in the third step of Figure 3. What information does it contain and how is it used by the system? \\
        6 & What happens when the LLM is queried for evaluation and rephrasing? Include information about the prompt and model settings. \\
        7 & In what format are heuristics injected into the LLM prompt after the designer selects guidelines? \\
        8 & How is the name field of the UI JSON created when the elements/groups in Figma are unnamed? \\
        9 & What does Figure 7 highlight about the types of NN Usability guidelines that GPT-4 is best equipped to handle? Describe the finding rather than listing data points. \\
        10 & Why was UI D in Figure 9 rated 1 for helpfulness? \\
    \bottomrule
    \end{tabular}
    \caption{Questions for the short-answer quiz.}
    \label{tab:app_eval_questions}
\end{table}

\section{Inter-annotator Agreement}

\begin{figure*}[h]
    \centering
    \includegraphics[width=0.5\linewidth]{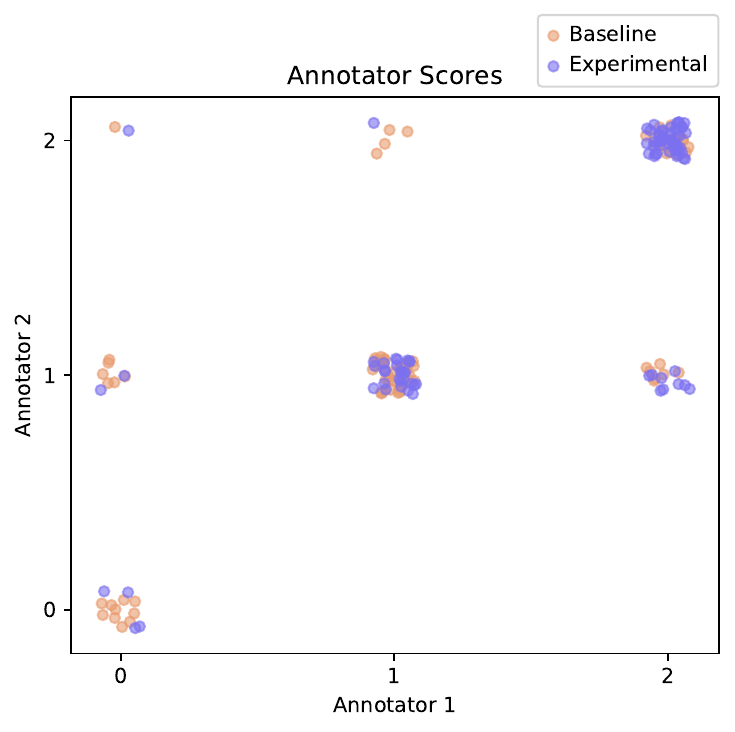}
    \caption{Response scores from two annotators. Two annotators assigned scores of 0, 1, or 2 to the short-answer quiz responses. Annotations show even distributions of baseline and experimental conditions with heavy clustering along the diagonal, indicating high agreement across conditions. These visual findings are supported by a Krippendorff's $\alpha$ of 0.75.}
    \label{fig:eval_iaa}
\end{figure*}

Figure~\ref{fig:eval_iaa} shows the distribution of scores for each annotator. It demonstrates heavy clustering along the diagonal without obvious clusters of baseline or experimental conditions, indicating agreement in general and across conditions. These results confirm that the quiz scores are of an acceptable quality for the remainder of our analysis.

\section{Evaluation Rubric}
\label{app:eval_rubric}

The annotators used the rubric in Table \ref{tab:app_eval_rubric} to assign ``quality'' scores based on correctness and completeness, as described in Section \ref{sec:eval_analysis}. Any responses containing factual errors received 0 credit. Responses containing no errors received full credit if they included all details from the rubric or partial credit if any detail was missing. Some responses indicated that participants had run out of time to answer them. The quality score and time to completion for these responses (5/180) were removed from analysis.

\clearpage
\FloatBarrier
\begin{longtable}{@{}c>{\raggedright}p{4.2cm}>{\raggedright}p{4.75cm}>{\raggedright}p{4.8cm}@{}}
\toprule
Q & Full credit details    & Full credit example & Partial credit example \tabularnewline
\midrule
\endfirsthead
\toprule
Q & Full credit details    & Full credit example & Partial credit example \tabularnewline 
\midrule
\endhead
1        & • Designer creates UI mockup. • Designer runs plugin. • System converts input to LLM.   & • The designer is building a Figma mockup which the plugin will convert to JSON, which will be used as input to the LLM. • Box A is the initial step where the designer builds a Figma mockup and the plugin converts it into a condensed JSON to be fed into the LLM. & • Box A represents the first phase of using the plugin where the designer prototypes the UI in Figma.            \tabularnewline
2 & • View feedback in an interpretable format. • AUTOMATIC 0 CREDIT if answer states that designer can update their design in this window. & • The Heuristic Evaluation Results Window shows guidelines violations as described by the LLM. The user will then use these results to update their design, and will rerun the evaluation once again. • It displays violations of the input heuristics to the user in the context of the UI prototype provided and allows users to also "X" out the feedback that they find unhelpful, which will be told to the LLM for future iterations. The designer can update their design based on this feedback and re-enter the feedback loop. & \tabularnewline
3 & • Designer can revise UI mockup based on feedback and then run the plugin again to get more feedback. Repetitive/iterative. & • The designer can continue to update their design and re-run the process of evaluation over and over. • used to update results and rerun & \tabularnewline
4 & • Feedback is hidden/dismissed. • Feedback added to prompt. • Prevent showing the same feedback again. & • The "X" button allows a user to dismiss feedback, which hides the feedback and adds this result to the LLM prompt for the next evaluation round. • The X button allows users to dismiss incorrect feedback and update the LLM prompt for future evaluations. & \tabularnewline
5 & • Contains semantic and visual information about the UI mockup. • Reduces number of tokens (for LLM context window). • Text input for LLM prompt. & • The condensed JSON representation removes unnecessary information and creates a structured format to send UI representations to LLMs in. By combining this structure with the guideline text, we can construct prompts for the LLMs to evaluate UI designs and also send back their recommendations in a similarly structured way. & • The ``Condensed JSON Representation'' allows the mockup from Figma to be converted to JSON such that a text-only LLM can capture visual elements of the UI as well as the textual content. \tabularnewline
6 & • Queried twice. • First query to identify violations in UI mockup according to selected heuristics. • Second query to convert violations into constructive suggestions. • Temperature = 0. • Prompt includes previously dismissed suggestions (optional to mention: ``self-reflection''). & • GPT-4 is called twice: once to evaluate guideline violations and another to rewrite the feedback into actionable advice. The prompt includes leveraging explicit instructions, setting the temperature to 0, and self-reflection. & • After identifying violations from the prompt containing the condensed JSON representation, the LLM rephrases the violations into constructive design advice. \tabularnewline
7 & • Numbered bullet points. • Partial credit for ``JSON'' or ``text.'' & &  \tabularnewline
8 & • LLM call. & • The name field is generated using an LLM call for all unnamed groups. • The label generation feature is executed through an LLM call. Prompt contains JSON data for all unlabeled features and the LLM creates a descriptive label for it. & \tabularnewline
9 & • Best on low-level details/worst on abstract details. • Partial credit for describing data points. & • It was strongest on clear, low-level heuristics like ``Consistency and Standards,'' ``Recognition rather than Recall,'' and ``Match Between System and Real World'' & • Figure 7 shows the accuracy and helpfulness of the model in terms of individual guidelines in the Performance Study. • best at consistency and standards • consistency and standards, recognition and recall, match between system and real world \tabularnewline
10 & • Feedback said that bounding boxes overlap but they do not overlap visually. • Weakness in text model. & • In UI D, the evaluators noted that this gallery group was incorrectly flagged for overlap because intersecting bounding boxes in the JSON misled GPT-4, demonstrating a core limitation of the text-only UI representation. & • rectangles are overlapping \tabularnewline
\bottomrule
\caption{Rubric for the short-answer quiz. The rubric provides all details required for full credit as well as a full-credit example for most questions. We provided a partial-credit examples upon request, which are also provided here. Annotators achieved Krippendorff's $\alpha = 0.75$, indicating moderate agreement.}
\label{tab:app_eval_rubric}
\end{longtable}

\section{Data Visualizations}

\begin{figure}
    \centering
    \includegraphics[width=\linewidth]{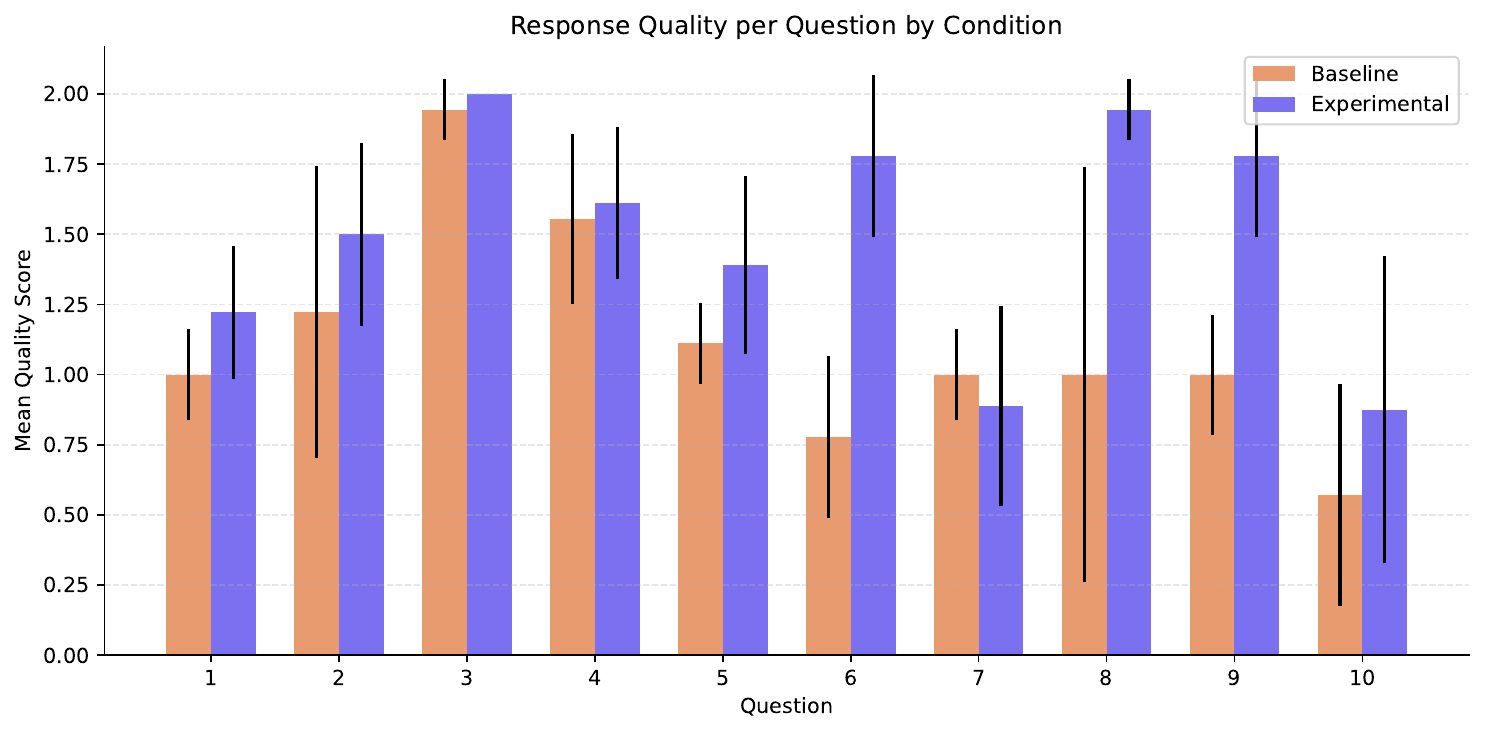}
    \caption{Response scores per question. Response quality was graded for correctness and completeness, ranging from 0 to 2. We report the mean score per question across conditions with 95\% confidence intervals. Improvements for questions 6 and 9 were statistically significant after Bonferroni correction according to Mann-Whitney U tests.}
    \label{fig:eval_perquestionacc}
\end{figure}

\begin{figure}
    \centering
    \includegraphics[width=\linewidth]{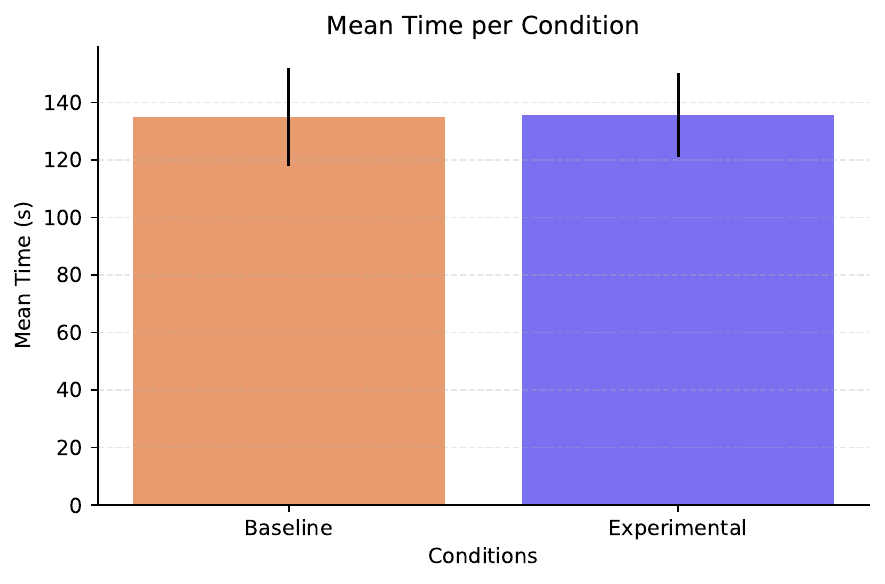}
    \caption{Mean time to completion across conditions with 95\% confidence intervals. Our analysis found no evidence of significant difference across conditions.}
    \label{fig:eval_time}
\end{figure}

\begin{figure}
    \centering
    \includegraphics[width=\linewidth]{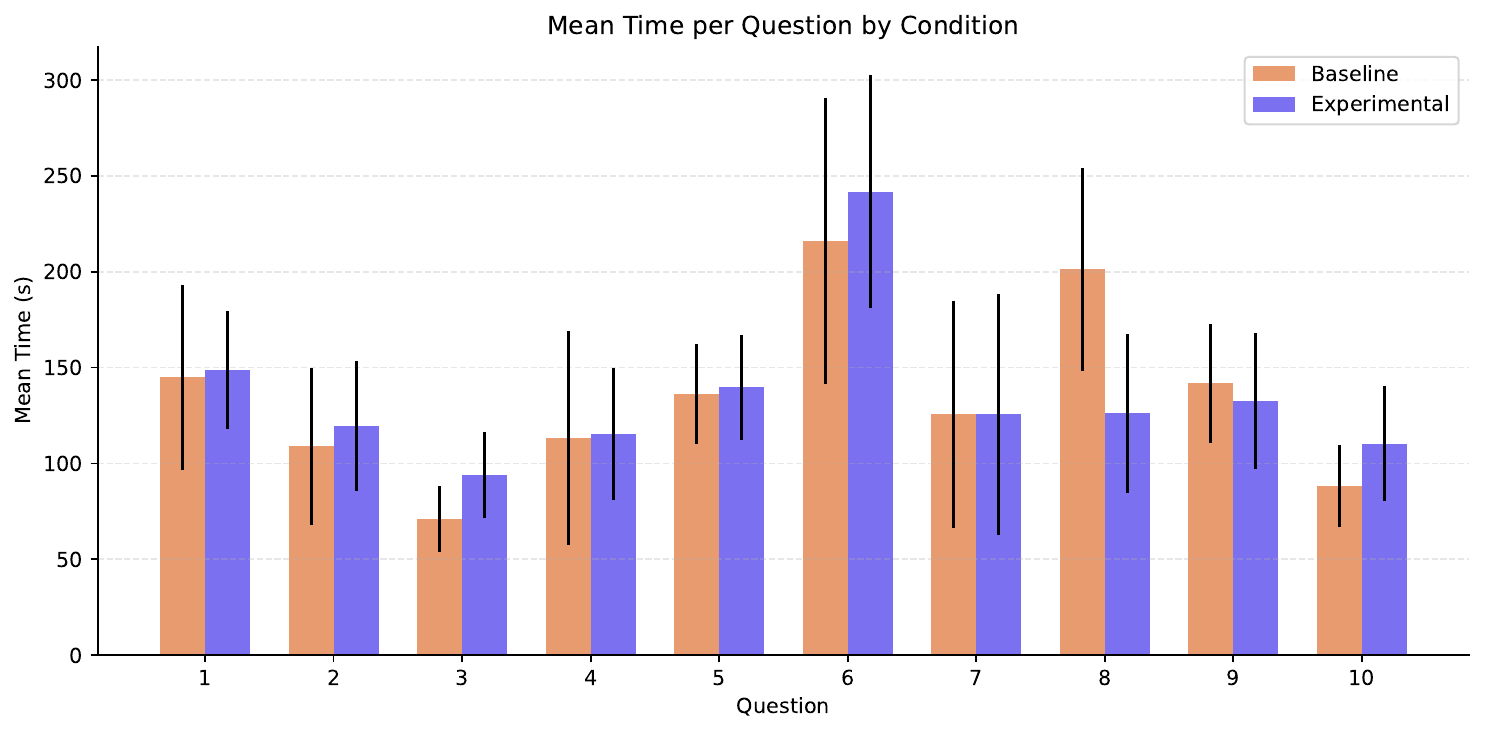}
    \caption{Mean time to completion per question across conditions with 95\% confidence intervals. Our analysis found no evidence of significant difference across conditions.}
    \label{fig:eval_perquestiontime}
\end{figure}

\begin{figure}
    \centering
    \includegraphics[width=\linewidth]{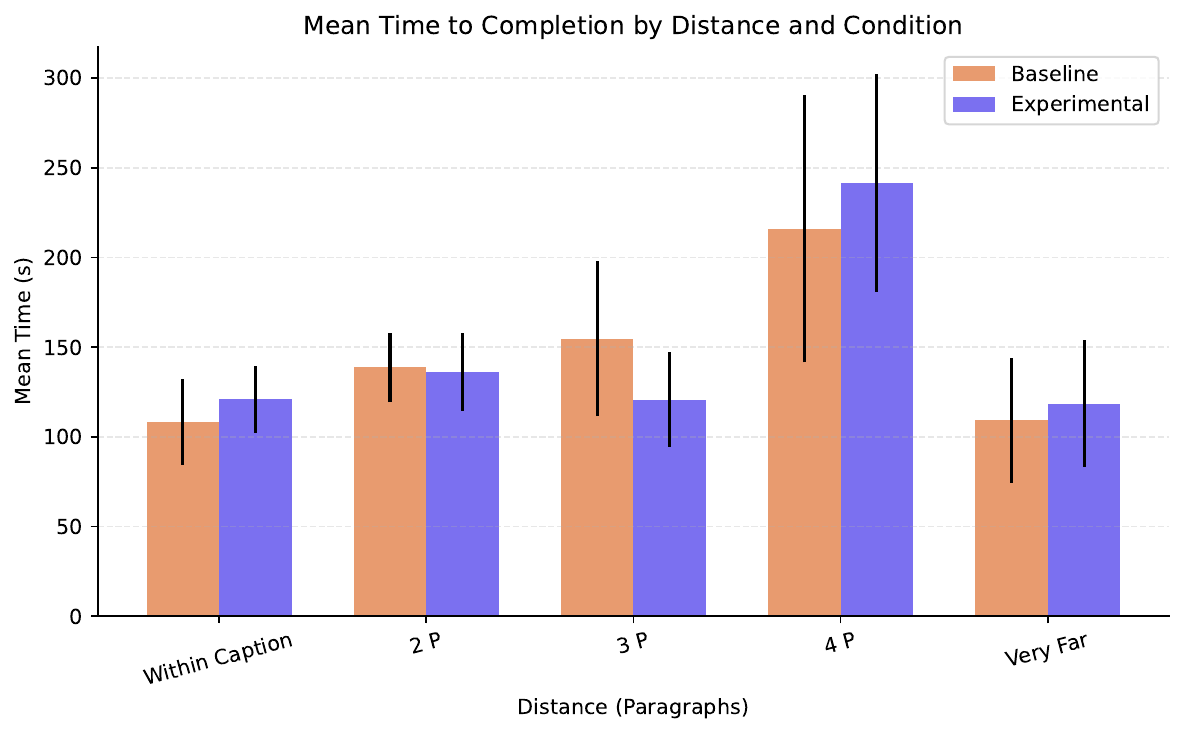}
    \caption{Mean time to completion by distance group across conditions with 95\% confidence intervals. Distance groups represent the number of paragraphs between details required for full credit.Our analysis found no evidence of significant difference across conditions for any distance group, even for groups with statistically significant (2P, 4P) improvement in response quality.}
\label{fig:eval_distancetime}
\end{figure}

\begin{figure}
    \centering
    \includegraphics[width=\linewidth]{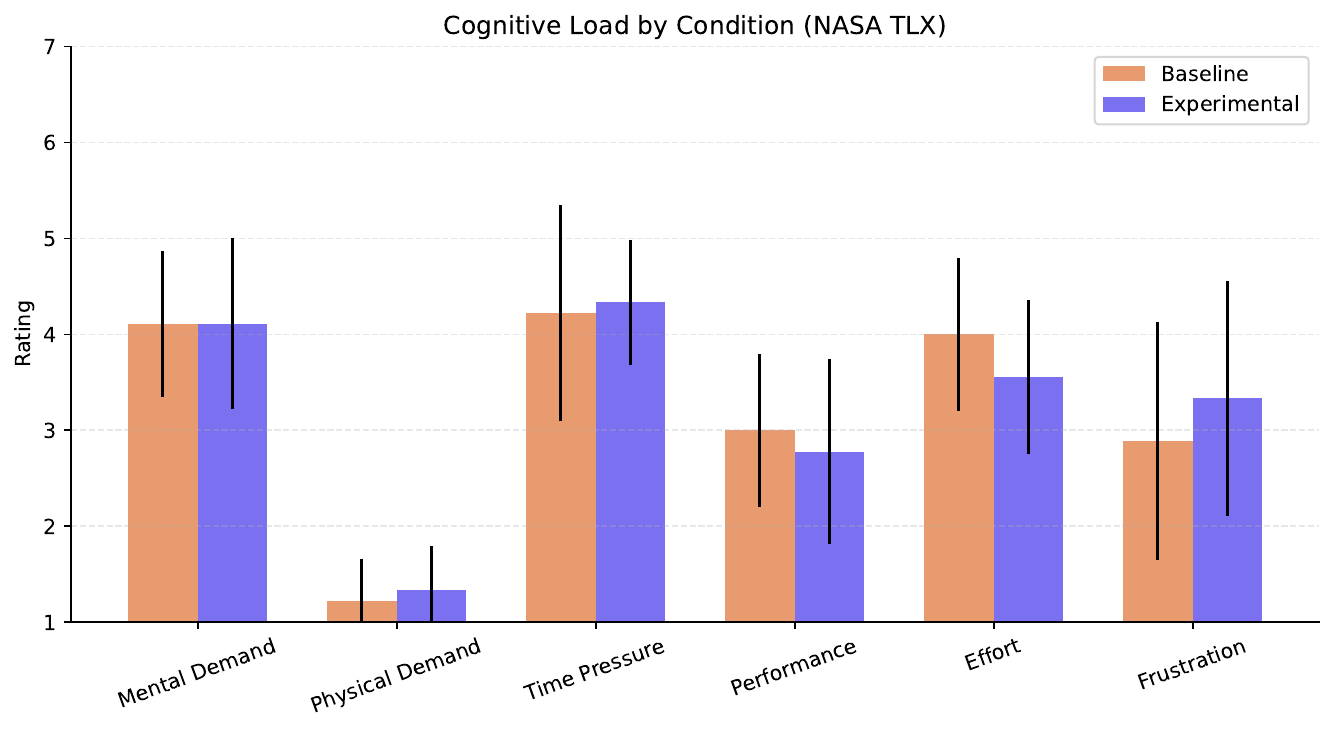}
    \caption{Perceived cognitive load across baseline and experimental groups. From left to right, participants reported mental demand, physical demand, time pressure, perceptions of their performance, effort, and frustration on a scale from 1 to 7. Despite some visual distances, the experimental group did not demonstrate statistically significant differences from baseline cognitive load.}
    \label{fig:eval_tlx}
\end{figure}

Additional data visualizations for the evaluation discussed in Section \ref{sec:eval} are shown in Figures \ref{fig:eval_perquestionacc}, \ref{fig:eval_time}, \ref{fig:eval_perquestiontime}, \ref{fig:eval_distancetime}, and \ref{fig:eval_tlx}.

\section{Statistical Tests}
\label{app:eval_statsig}

We report the following additional statistical tests for our analysis in Section \ref{sec:findings} (see Tables \ref{tab:eval_questions}, \ref{tab:eval_questions_distance}, \ref{tab:eval_time}, \ref{tab:eval_time_distance}, \ref{tab:app_eval_time_equiv}, and \ref{tab:app_eval_tlx_equiv}).

\begin{table}
\centering
\begin{tabular}{ccccl}
\toprule
Question & $U$ & $p$ & $r$ & $p_b$    \\
\midrule
1  & 53   & 0.089   & -0.309 &        \\
2  & 47   & 0.288   & -0.160 &        \\
3  & 45   & 0.187   & -0.111 &        \\
4  & 43   & 0.425   & -0.062 &        \\
5  & 52.5 & 0.112   & -0.296 &        \\
6  & 75   & 0.00079 & -0.852 & \textbf{0.0079} \\
7  & 32.5 & 0.815   & 0.198  &        \\
8  & 52   & 0.0309  & -0.444 & 0.309  \\
9  & 56   & 0.00272 & -0.778 & \textbf{0.0272} \\
10 & 35.5 & 0.193   & -0.268 &        \\
\bottomrule
\end{tabular}
\caption{Mann-Whitney U tests for response quality per question. Improvements in questions 6 and 9 were statistically significant after Bonferroni correction ($p_b$).}
\label{tab:eval_questions}
\end{table}

\begin{table}
\centering
\begin{tabular}{ccccl}
\toprule
Distance & $U$ & $p$ & $r$ & $p_b$ \\
\midrule
Within caption & 418   & 0.161   & -0.407 & \\
2 paragraphs   & 223.5 & 0.00115 & -0.552 & \textbf{000575} \\
3 paragraphs   & 199   & 0.0471  & -0.301 & 0.2355   \\
4 paragraphs   & 75    & 0.00079 & -0.852 & \textbf{0.00395} \\
Very far       & 142.5 & 0.406   & -0.048 & \\
\bottomrule
\end{tabular}
\caption{Mann-Whitney U tests for response quality by distance group. Improvement in quality was statistically significant for 2P and 4P after Bonferroni correction ($p_b$).}
\label{tab:eval_questions_distance}
\end{table}

\begin{table}
\centering
\begin{tabular}{cccl}
\toprule
Question & $U$    & $p$               \\
\midrule
1        & 34.0   & 0.596            \\
2        & 35.0   & 0.659            \\
3        & 23.0   & 0.133            \\
4        & 30.0   & 0.377            \\
5        & 34.0 & 0.596            \\
6        & 34.0   & 0.596           \\
7        & 43.0 & 0.860            \\
8        & 52.0   & 0.139          \\
9        & 35.0   & 0.758           \\
10       & 21.0 & 0.463            \\
\bottomrule
\end{tabular}
\caption{Mann-Whitney U tests for time to completion per question.}
\label{tab:eval_time}
\end{table}

\begin{table}
\centering
\begin{tabular}{cccl}
\toprule
Distance & $U$    & $p$               \\
\midrule
Within caption        & 294.0   & 0.226            \\
2 paragraphs        & 177.0   & 0.438            \\
3 paragraphs       & 145.0   & 0.986            \\
4 paragraphs       & 34.0   & 0.596            \\
Very far       & 130.0 & 0.843            \\
\bottomrule
\end{tabular}
\caption{Mann-Whitney U tests for time to completion by distance group.}
\label{tab:eval_time_distance}
\end{table}

\begin{table}
\centering
\begin{tabular}{lc}
\toprule
Condition & $p$ \\
\midrule
Overall & 0.688 \\
Question 1 & 0.294 \\
Question 2 & 0.365 \\
Question 3 & 0.579 \\
Question 4 & 0.299 \\
Question 5 & 0.202 \\
Question 6 & 0.545 \\
Question 7 & 0.328 \\
Question 8 & 0.938 \\
Question 9 & 0.334 \\
Question 10 & 0.677 \\
Within caption & 0.313 \\
2 paragraphs & 0.125 \\
3 paragraphs & 0.707 \\
4 paragraphs & 0.545 \\
Very far & 0.379 \\
\bottomrule
\end{tabular}
\caption{TOST equivalence tests for time to completion.}
\label{tab:app_eval_time_equiv}
\end{table}

\begin{table}
\centering
\begin{tabular}{lc}
\toprule
Condition & $p$ \\
\midrule
Mental demand & 0.208 \\
Physical demand & 0.124 \\
Time pressure & 0.283 \\
Performance & 0.335 \\
Effort & 0.462 \\
Frustration & 0.475 \\
\bottomrule
\end{tabular}
\caption{TOST equivalence tests for cognitive load (NASA Task Load Index).}
\label{tab:app_eval_tlx_equiv}
\end{table}

\end{document}